\documentclass[conference,compsoc]{IEEEtran}
\IEEEoverridecommandlockouts

\usepackage{graphicx}
\usepackage{amsmath}
\usepackage{amssymb}
\usepackage{amsthm}
\usepackage{booktabs}
\usepackage{mathtools}
\usepackage{algorithm}
\usepackage{algpseudocode}
\usepackage[table]{xcolor}
\usepackage{framed}
\usepackage{ragged2e}
\usepackage{multirow}
\usepackage{capt-of}

\newenvironment{assurancecase}{%
	\def\FrameCommand##1{%
		{\color{black!55}\vrule width 0.5pt}%
		{\setlength{\fboxsep}{5pt}\colorbox{black!5}{##1}}%
	}%
	\MakeFramed{\advance\hsize-\width\FrameRestore}%
}{%
	\endMakeFramed
}

\newcommand{\ShadedComment}[1]{%
	\Statex
	\hspace*{-\leftmargin}%
	\colorbox{gray!10}{%
		\makebox[\dimexpr\linewidth+\leftmargin-2\fboxsep\relax][r]{%
			{\normalsize #1}%
		}%
	}%
}

\usepackage{fontawesome5}
\usepackage{tabularx}
\usepackage{array}

\newcolumntype{Y}{>{\raggedright\arraybackslash}X}
\newcolumntype{C}{>{\centering\arraybackslash}X}

\newtheorem{theorem}{Theorem}[section]
\newtheorem{lemma}[theorem]{Lemma}

\newtheorem{proposition}[theorem]{Proposition}

\usepackage[hidelinks]{hyperref}

\begin{document}
\title{Where Did My Tokens Go? Verifiable Token Billing for LLM Services}

\author{%
	\begin{tabular}{@{}c@{\hspace{0.04\textwidth}}c@{\hspace{0.04\textwidth}}c@{}}
		\parbox[t]{0.29\textwidth}{\centering\small
			{\large Jiayuan Du}\\
			Beijing University of Posts and\\
			Telecommunications\\
			Beijing, China\\
			jydu0515@bupt.edu.cn}
		&
		\parbox[t]{0.29\textwidth}{\centering\small
			{\large Yebo Feng\textsuperscript{*}}\\
			Nanyang Technological University\\
			Singapore\\
			yebo.feng@ntu.edu.sg}
		&
		\parbox[t]{0.29\textwidth}{\centering\small
			{\large Lin Zhang}\\
			Beijing University of Posts and\\
			Telecommunications\\
			Beijing Big Data Center\\
			Beijing, China\\
			zhanglin@bupt.edu.cn}
		\\[0.8em]
		&
		\parbox[t]{0.29\textwidth}{\centering\small
			{\large Konglin Zhu\textsuperscript{*}}\\
			Beijing University of Posts and\\
			Telecommunications\\
			Beijing Key Laboratory of\\
			Multimodal Data Intelligent\\
			Perception and Governance\\
			Beijing, China\\
			klzhu@bupt.edu.cn}
		&
	\end{tabular}%
	\thanks{Corresponding authors: Yebo Feng and Konglin Zhu.}%
}

\maketitle
		
\begin{abstract}
	Users of large language model (LLM) services receive token bills
	with limited evidence of which calls incurred their charges.
	In broker-mediated agent workflows, calls arise after authorization
	and billing evidence spans service domains. A bill can count one
	genuine receipt twice and omit another. Valid metering evidence
	can also be attributed to the wrong execution.
	We present \textsc{TokenProv}, which uses authenticated provider
	acceptances to establish settlement coverage and verify token
	charges. An authenticated causal history links runtime attempts
	to the signed policy, and jointly signed handoffs bind intent,
	dispatch, and acceptance to the same call. Handoffs submitted
	independently of the bill can reveal omitted calls. The auditor
	requires a one-to-one match between observable acceptances and
	settled receipts. For each acceptance, a common provider relation binds metered usage to the cache-dependent execution of the accepted input using the registered model and serving rules, and checks the
	charge under the pricing rule.
	To limit serving-path overhead, providers can use attested
	state transitions in trusted execution environments or defer
	proof generation until audit by committing to witnesses in
	receipts before settlement. The auditor recomputes the bill from the verified charges. We prove audit soundness relative to authenticated evidence.
	Across five models, \textsc{TokenProv} with signed provider receipts
	achieves 94.4--107.4\% of baseline throughput with 54--66\,ms of online non-LLM processing per request. In a separate eight-call replay, deferring scoped proofs reduces foreground protocol time from
	1.20\,s to 8.65\,ms (99.28\%). The auditor rejects all 1,600 validly
	signed attack queries across 100 adaptive episodes.
\end{abstract}

\IEEEpeerreviewmaketitle

\section{Introduction}
\label{sec:introduction}
\begin{figure}[t]
	\centering
	\includegraphics[width=\linewidth]{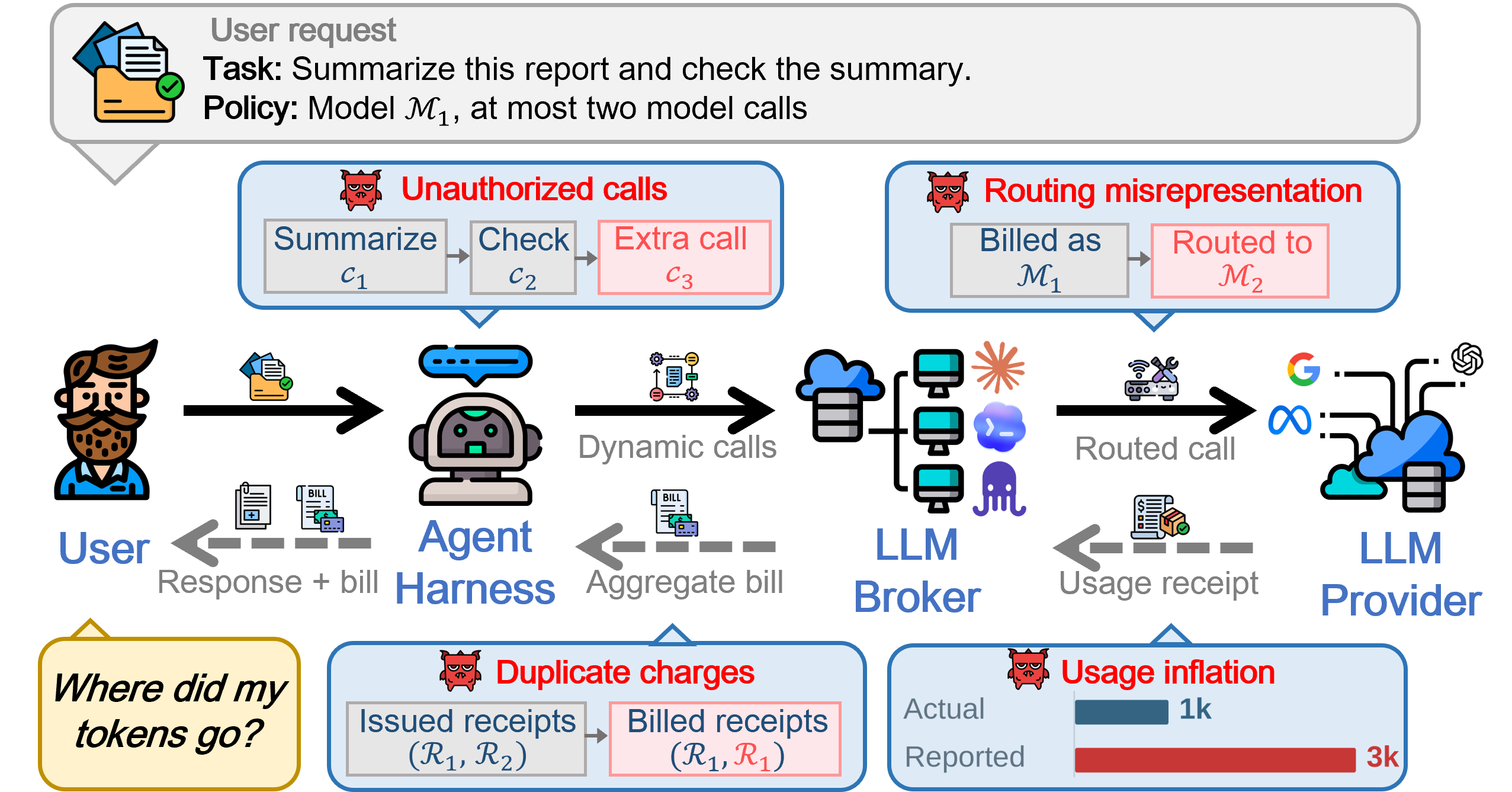}
	\caption{Billing integrity risks in a broker-mediated agent workflow.
		The token-inflation example follows prior work
		~\cite{hoque2026tokeninflation}.}
	\label{fig:motivation}
\end{figure}

Large language model (LLM) agents combine model calls with tool
interactions, and intermediate results can determine later
calls~\cite{debenedetti2024agentdojo}.
In broker-mediated services, the agent harness generates calls,
the broker selects a provider deployment~\cite{ong2025routellm},
and the provider executes the calls and reports token usage.
The user receives a response and an aggregate bill that can include
charges for hidden operations~\cite{sun2025invisible}.
In Figure~\ref{fig:motivation}, the user authorizes at most two calls
to $\mathcal{M}_1$ to draft and check a report summary.
The harness can add a third call. The broker can bill for
$\mathcal{M}_1$ while routing to a cheaper model $\mathcal{M}_2$,
or reuse a receipt to bill the same call twice. The provider can
inflate reported token usage
~\cite{artola2025overcharging,hoque2026tokeninflation}.
A plausible summary can thus accompany charges for unauthorized
or misrepresented work.

At the harness, prior work traces agent workflows and scopes
delegated authority~\cite{souza2025provagent,south2025delegation}.
At the broker, GateScope tests for model substitution and billing
discrepancies, and the Attested Gateway Runtime signs routing
evidence~\cite{lin2026gateways,wang2026gateway}.
At the provider, existing work verifies inference, audits hidden
reasoning tokens, or tests usage reports across
queries~\cite{qu2025zkgpt,sun2025coin,velasco2026auditing}.
Even when two authorized calls are executed and metered correctly,
a broker can replace one genuine receipt with a copy of the other.
A verifier that checks each receipt independently and recomputes
the total from the listed charges can accept the resulting bill,
which charges for one call twice and omits the other.

Verifying the final bill requires reconciling it with the
observable call history rooted in the user's signed authorization.
\par\smallskip
\begingroup
\setlength{\fboxsep}{6pt}
\noindent\colorbox{black!5}{%
	\parbox{\dimexpr\linewidth-2\fboxsep\relax}{%
		\setlength{\parindent}{0pt}
		\justifying
		\noindent\textit{How can we establish \textbf{end-to-end verifiable
				token-spending provenance} from signed user authorization to
			request-level settlement when model calls arise at runtime
			and cross mutually distrustful service domains?}%
	}%
}\par
\endgroup
\smallskip

\noindent
\emph{Runtime coverage.}
Branching and retries create call attempts after authorization.
The audit must check observable attempts against the signed policy
and match observable provider acceptances to settled receipts
one-to-one.
\emph{Cross-domain binding.}
The broker can transform inputs and choose deployments.
Verification must bind intent, dispatch, and provider acceptance
to the same call.
\emph{Execution-consistent accounting.}
Individually valid execution and metering records can refer to
different invocations or cache states. The protocol must bind usage
to the accepted execution and verify the charge under the pricing
rule fixed at dispatch, while limiting serving-path overhead.

We present \textsc{TokenProv}, a protocol for verifiable
request-level billing in LLM services.

\noindent\textbf{Technical highlights.}
\emph{Authenticated lifecycle and exact settlement.}
An authenticated causal history links runtime call attempts to
signed user authorization. The audit matches observable provider
acceptances to settled receipts one-to-one and recomputes the
billed total.
\emph{Bilateral handoffs.}
Bilateral handoffs bind the harness's intent, the broker's dispatch,
and the provider's acceptance to the same call. A holder can submit
a retained broker--provider handoff directly to the auditor. This
lets the auditor detect an omitted receipt even when the broker's
invoice is internally consistent.
\emph{Unified provider assurance.}
A common relation binds the accepted input, deployment,
cache-dependent execution, metered usage, and charge to the same
invocation. The auditor verifies this relation through a TEE-attested
state transition or a cryptographic proof. In the proof-based path,
the provider commits to the witness in its receipt before settlement.
Proof generation can then be deferred until audit time.

\noindent\textbf{Evaluation highlights.}
Across five models, \textsc{TokenProv} with signed provider receipts
achieves 94.4--107.4\% of \textsc{Vanilla}'s mean throughput.
Online non-LLM processing takes 54--66\,ms per request.
In a separate eight-call replay, deferring scoped proofs reduces
foreground protocol time from 1.20\,s to 8.65\,ms (99.28\%).
The auditor rejects all 1,600 tested adaptive attack queries with
valid signatures.

\noindent\textbf{Contributions.}
\begin{itemize}
	\item We formalize end-to-end token-spending provenance for
	dynamic LLM workflows and define request-level billing
	correctness relative to authenticated evidence.
	
	\item We design \textsc{TokenProv} to compose authenticated call
	histories, bilateral handoffs, and provider assurance. The audit
	links each settled token charge to its authorized execution and
	matches observable provider acceptances to settled receipts
	one-to-one. A common provider relation supports TEE-attested
	state transitions and deferred proofs over committed witnesses.
	
	\item We prove provider-assurance soundness, settlement exactness,
	and end-to-end audit soundness. Except with negligible probability,
	an accepting audit admits a joint witness for the full provenance
	relation. We establish audit confidentiality under an explicit
	leakage model. For reproducible integrity failures, the reported
	responsibility interval covers every fault location consistent with
	the retained evidence.
	
	\item We implement \textsc{TokenProv} and compare serving latency
	and throughput against three baselines across five LLMs.
	We measure scaling with call and lifecycle-event counts and
	quantify the costs of scoped provider proofs, evidence retention,
	and ledger-backed settlement. Adaptive attacks with valid
	signatures, evidence-splicing tests, and collusion controls
	assess audit effectiveness. Fault injections evaluate
	localization, and ablations quantify the effects of individual
	audit checks.
\end{itemize}

\noindent\textbf{Artifacts.}
The prototype and evaluation scripts are available at
\faGithub\,\url{https://github.com/Dujiayuan-create/TokenProv}.

\section{Background and Preliminaries}
\label{sec:preliminaries}

\subsection{Notation and Cryptographic Primitives}
\noindent\textbf{Notation.}
Let $\lambda$ denote the security parameter and
$\mathsf{negl}(\lambda)$ a negligible function. We write
$[n]\triangleq\{1,\ldots,n\}$, $\parallel$ for concatenation,
$\epsilon$ for the empty sequence, $\bot$ for an undefined value,
and $\mathbf{1}[\cdot]$ for the indicator function.
For an ordered record sequence $\mathbf{X}$,
$\mathsf{MRoot}(\mathbf{X})$ denotes its Merkle
root~\cite{merkle1988digital}. We write $a\prec_{\mathsf{prot}}b$
when the authenticated contents of $b$ bind a reference to $a$.
Appendix~\ref{app:notation-summary} summarizes the
protocol-specific notation.

\noindent\textbf{Cryptographic primitives.}
We use a collision-resistant hash $\mathsf{Hash}$ on byte strings.
For structured $z$, $\mathsf{Hash}(z)$ abbreviates
$\mathsf{Hash}(\mathsf{Encode}(z))$, where $\mathsf{Encode}$ is
deterministic and injective.
We use an EUF-CMA-secure signature scheme~\cite{goldwasser1988digital}.
For signer $E$ with key pair $(pk_E,sk_E)$, write
$\mathsf{Sig}_E(z)\triangleq\mathsf{Sign}_{sk_E}(\mathsf{Hash}(z))$.
A signature $\sigma$ is valid when
$\mathsf{VerifySig}_{pk_E}(\mathsf{Hash}(z),\sigma)=1$.
Multiple arguments to $\mathsf{Hash}$ or $\mathsf{Sig}_E$ denote
a single ordered tuple.
For a computationally hiding and binding commitment scheme with
public parameters $pp$, $(C,d)\leftarrow\mathsf{Commit}(pp,z)$
returns a commitment and its opening information.
A valid opening satisfies $\mathsf{Open}(pp,C,z,d)=1$.

\subsection{Commit-and-Prove Arguments}
\label{sec:commit-and-prove}
A commit-and-prove argument
$\Pi_{\mathsf{CP}}$~\cite{campanelli2019legosnark} for an NP
relation $\mathcal{R}$ takes public input $(x,C_w)$ and private
input $(w,d_w)$. Here $C_w$ is a commitment to $w$ under public
parameters $pp$, with opening $d_w$. The argument verifies
\begin{equation}
	\mathcal{R}_{\mathsf{CP}}
	\triangleq
	[\mathcal{R}(x,w)=1]
	\land [\mathsf{Open}(pp,C_w,w,d_w)=1].
	\label{eq:cp-relation}
\end{equation}
The prover computes $\pi\leftarrow
\mathsf{Prove}(pp,x,C_w;w,d_w)$. The verifier accepts when
$\mathsf{Verify}(pp,x,C_w,\pi)=1$.
We assume completeness, zero knowledge, and knowledge soundness
with joint extraction for adaptively submitted proofs in one audit.

\subsection{Attested State Transitions}
\label{sec:attested-execution}
\noindent\textbf{Executor attestation.}
We use remote attestation to authenticate a protected
executor~\cite{volos2018graviton,ivanov2023sage}.
Let $\mu$ be the approved measurement of $\mathsf{Prog}$ and
$d_D\triangleq\mathsf{Hash}(D)$ the digest of the registered
deployment descriptor $D$. The executor holds a key pair
$(vk_T,sk_T)$ and keeps $sk_T$ within the protected boundary.
At initialization, the verifier issues a fresh challenge $\nu_T$.
It checks that the quote $Q$ binds $vk_T$ to $\mu$ and $d_D$
under this challenge by requiring
$\mathsf{VerifyQuote}(Q;\mu,d_D,vk_T,\nu_T)=1$.

\noindent\textbf{Transition verification.}
For input $x_j$ and predecessor state $S_j$, the measured program
computes
\begin{equation}
	(y_j,S_{j+1})\leftarrow F_{\mathsf{Prog},D}(S_j,x_j),
	\label{eq:protected-state-transition}
\end{equation}
where $y_j$ is the output and $S_{j+1}$ the successor state.
Let $h_Q\triangleq\mathsf{Hash}(Q)$, $s_k\triangleq\mathsf{Hash}(S_k)$,
and $\mathsf{DST}_{\mathsf{att}}$ be the domain-separation tag.
The executor forms and signs the transition record
\[
z_j^T\triangleq
(\mathsf{DST}_{\mathsf{att}},h_Q,\mu,d_D,x_j,y_j,s_j,s_{j+1}),
\sigma_j^T\leftarrow\mathsf{Sig}_T(z_j^T).
\]
A rollback-resistant continuity mechanism authenticates the
admissible predecessor digest $\bar{s}_j$ for transition
$j$~\cite{parno2011memoir}. The auditor validates $Q$ against the
authenticated initialization context, including the original
challenge $\nu_T$. It reconstructs $z_j^T$ from that context,
the input and output under verification, and the supplied state
digests. It then checks
\begin{equation}
	[\mathsf{VerifySig}_{vk_T}(\mathsf{Hash}(z_j^T),\sigma_j^T)=1]
	\land [s_j=\bar{s}_j].
	\label{eq:attested-transition-check}
\end{equation}

\section{Formalizing Token-Spending Provenance}
\label{sec:model}
Given authenticated evidence, an auditor checks whether the
observable call history complies with the signed authorization
and substantiates the bill retained by the user.

\subsection{Serving and Accounting}
\label{sec:serving-semantics}
\noindent\textbf{Service and audit roles.}
The user $U$ issues signed authorization $\mathsf{Auth}_r$
for request $r$. The agent harness $H$ generates call attempts
whose identifiers form $\mathcal{Q}_r$. For each dispatched
attempt, the broker $B$ selects a provider deployment and serving
configuration. The provider $P$ reports the outcome, metered
usage, and charge for each accepted call. Its accepted call
identifiers form $\mathcal{Q}_r^P\subseteq\mathcal{Q}_r$.
The broker reports the request charge $B_r$.
The append-only ledger $L$ anchors deployment registrations
and finalized settlements. The off-chain repository $V$ retains
private payloads and commitment openings.
Figure~\ref{fig:tokenprov-workflow} shows the evidence flow.

\begin{figure}[t]
	\centering
	\includegraphics[width=0.9\linewidth]{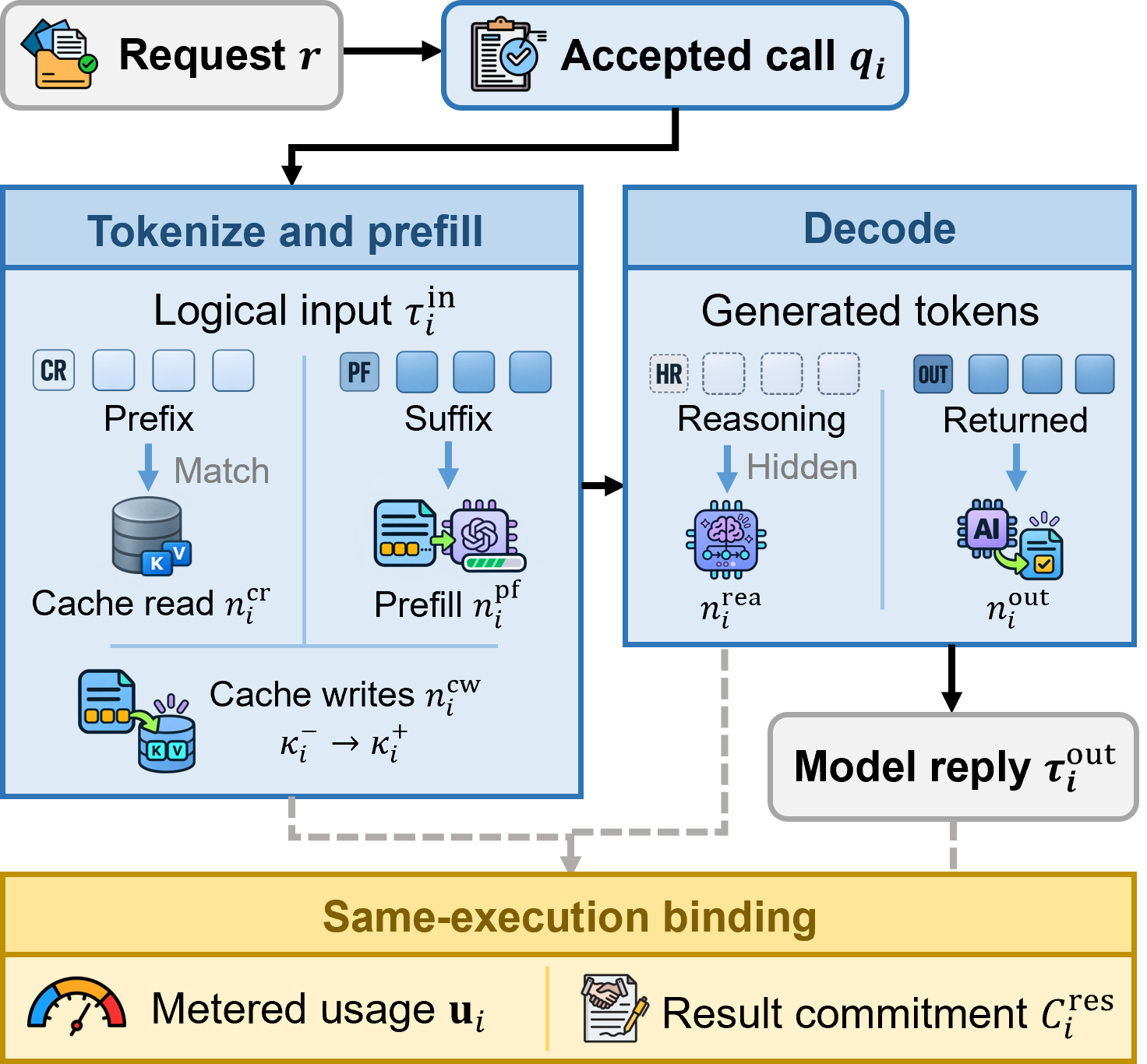}
	\caption{Token metering for an accepted model call. Provider
		assurance binds metered usage $\mathbf{u}_i$ and the result
		commitment $C_i^{\mathsf{res}}$ to the same execution.}
	\label{fig:provider-token-flow}
\end{figure}

\noindent\textbf{Provider input.}
For attempt $i$, the harness assigns a fresh identifier
$\mathsf{cid}_i$ and forms the intent
\begin{equation}
	\iota_i\triangleq(r,\mathsf{cid}_i,\mathbf{m}_i,\vartheta_i),
	\label{eq:call-intent}
\end{equation}
where $\mathbf{m}_i$ is the structured message sequence and
$\vartheta_i$ specifies the call's execution constraints.
For a dispatched attempt, the broker selects the configuration
\[
\theta_i\triangleq
\bigl(\mathsf{pid}_i,\mathsf{mid}_i,\mathsf{ep}_i,
\theta_i^{\mathsf{dec}},\theta_i^{\mathsf{cache}},
\theta_i^{\mathsf{meter}},\mathsf{pv}_i\bigr).
\]
The fields specify the provider, model, endpoint, decoding settings,
cache policy, metering rules, and pricing version, respectively.
The broker applies transformation rule $\rho_i$ to form
\begin{equation}
	\mathbf{m}'_i\triangleq\mathsf{Transform}_{\rho_i}(\mathbf{m}_i),
	\quad q_i\triangleq(r,\mathsf{cid}_i,\mathbf{m}'_i,\theta_i).
	\label{eq:provider-call}
\end{equation}
The bilateral handoffs bind $\iota_i$ and $q_i$ to the same request
and call identifiers. The broker--provider handoff also binds
$\rho_i$ and the deployment registration
(Section~\ref{sec:authenticated-lifecycle}).
For each accepted call $\mathsf{cid}_i\in\mathcal{Q}_r^P$,
the registered canonicalizer and tokenizer yield
\begin{equation}
	\boldsymbol{\tau}_i^{\mathsf{in}}\triangleq
	\mathsf{Tok}_{\theta_i}
	\bigl(\mathsf{Canon}_{\theta_i}(\mathbf{m}'_i)\bigr).
	\label{eq:provider-input}
\end{equation}
The logical input includes any prefix whose KV state is reused.
Figure~\ref{fig:provider-token-flow} shows where metering occurs
during the execution of the accepted call.

\noindent\textbf{Cache-dependent execution.}
Given a logical KV-cache pre-state $\kappa_i^{-}$ and private
randomness $\xi_i$, the registered inference procedure produces
\begin{equation}
	\bigl(g_i,\boldsymbol{\tau}_i^{\mathsf{out}},\phi_i,\kappa_i^{+}\bigr)
	\leftarrow
	\mathsf{Infer}_{\theta_i}
	\bigl(\boldsymbol{\tau}_i^{\mathsf{in}},\kappa_i^{-};\xi_i\bigr),
	\label{eq:provider-execution}
\end{equation}
where $g_i$ is the execution trace, $\boldsymbol{\tau}_i^{\mathsf{out}}$
the returned tokens, $\phi_i$ the termination code, and $\kappa_i^{+}$
the cache post-state. Let $\boldsymbol{\tau}_i^{\mathsf{cr}}$ be the
prefix whose KV state is reused~\cite{zheng2024sglang} and
$\boldsymbol{\tau}_i^{\mathsf{pf}}$ the suffix assigned to
prefill~\cite{agrawal2024sarathi}. The trace records the reused
prefix, completed prefill positions, and the phase of each
generated token. The input partition and cache update satisfy
\begin{equation}
	\begin{gathered}
		\boldsymbol{\tau}_i^{\mathsf{in}}
		=
		\boldsymbol{\tau}_i^{\mathsf{cr}}
		\parallel\boldsymbol{\tau}_i^{\mathsf{pf}},\\
		\mathsf{CacheStep}_{\theta_i^{\mathsf{cache}}}
		\bigl(\kappa_i^{-},\boldsymbol{\tau}_i^{\mathsf{in}},
		\boldsymbol{\tau}_i^{\mathsf{cr}},\kappa_i^{+}\bigr)=1.
	\end{gathered}
	\label{eq:logical-cache-step}
\end{equation}
A hit requires a pre-state entry matching the prefix, deployment,
and cache-rule version, and a miss gives
$\boldsymbol{\tau}_i^{\mathsf{cr}}=\epsilon$.

\noindent\textbf{Token usage and charges.}
For each accepted call, define the usage vector
\begin{equation}
	\mathbf{u}_i\triangleq
	\bigl(n_i^{\mathsf{pf}},n_i^{\mathsf{cr}},n_i^{\mathsf{cw}},
	n_i^{\mathsf{rea}},n_i^{\mathsf{out}}\bigr).
	\label{eq:call-usage}
\end{equation}
The prefill count $n_i^{\mathsf{pf}}$ counts completed prefill
positions in $g_i$. The counts satisfy
\[
n_i^{\mathsf{pf}}\leq|\boldsymbol{\tau}_i^{\mathsf{pf}}|,
\quad n_i^{\mathsf{cr}}=|\boldsymbol{\tau}_i^{\mathsf{cr}}|,
\quad n_i^{\mathsf{out}}=|\boldsymbol{\tau}_i^{\mathsf{out}}|.
\]
The prefill bound holds with equality when the suffix is fully
processed.
The cache-write count
$n_i^{\mathsf{cw}}=\mathsf{CacheWrite}_{\theta_i^{\mathsf{cache}}}
(\kappa_i^{-},\kappa_i^{+})$
sums the token lengths of entries newly present in the post-state
under the registered cache policy. With KV caching disabled,
$n_i^{\mathsf{cr}}=n_i^{\mathsf{cw}}=0$.
The reasoning count
$n_i^{\mathsf{rea}}=\mathsf{ReasonMeter}_{\theta_i^{\mathsf{meter}}}(g_i)$
counts generated positions classified as reasoning, which may include
hidden reasoning tokens~\cite{sun2025coin}.
The authenticated $B\rightarrow P$ handoff fixes the pricing version
$\mathsf{pv}_i$ before execution. The call charge is
\begin{equation}
	b_i\triangleq\mathsf{Price}_{\mathsf{pv}_i}(\theta_i,\mathbf{u}_i).
	\label{eq:call-charge}
\end{equation}
The pricing rule specifies how to charge overlapping usage
categories and returns an integer amount in the specified currency
unit using fixed rounding. A response served directly from cache incurs zero new model-token usage.
An authenticated reuse fee may enter settlement as an auxiliary fee
if the request policy permits it.

\subsection{Request-Level Audit Model}
\label{sec:authorization-boundary}
\noindent\textbf{Request authorization.}
For request $r$, let $C_r^{\mathsf{in}}$ be the user's commitment
to the private input, $\Psi_r$ the execution and billing policy,
and $\nu_r$ a fresh nonce. The user $U$ issues
\begin{equation}
	\begin{aligned}
		\mathsf{auth}_r &\triangleq (r,C_r^{\mathsf{in}},\Psi_r,\nu_r),\\
		\mathsf{Auth}_r &\triangleq
		\bigl(\mathsf{auth}_r,\mathsf{Sig}_U(\mathsf{auth}_r)\bigr).
	\end{aligned}
	\label{eq:request-authorization}
\end{equation}
$\mathsf{VerifyAuth}(\mathsf{Auth}_r)$ verifies the signature
under $pk_U$ and checks the signed request identifier against $r$.
The policy $\Psi_r$ specifies permitted providers, model artifact
digests, transformations, and assurance profiles, together with
metering and pricing rules. It sets limits on token usage, attempts,
retries, and spending. It also specifies billable outcomes, permitted
auxiliary fees, evidence retention, and audit deadlines.
With domain-separation tag $\mathsf{DST}_{\mathsf{auth}}$, define
\begin{equation}
	h_r^{\mathsf{auth}}\triangleq
	\mathsf{Hash}\bigl(\mathsf{DST}_{\mathsf{auth}},pk_U,\mathsf{auth}_r\bigr).
	\label{eq:authorization-digest}
\end{equation}
Lifecycle, handoff, and settlement records bind to the same
request context $(r,h_r^{\mathsf{auth}})$.

\noindent\textbf{Observable calls.}
Fix $(r,h_r^{\mathsf{auth}})$. The audit transcript $\mathcal{T}_r$
contains records opened against finalized request commitments.
It includes independently retained boundary records submitted
by the audit cutoff with valid signatures and request bindings.
The supplied private payloads and openings form $\mathcal{W}_r$.
The audit is incomplete if required evidence is missing at the response deadline in $\Psi_r$.

$\mathsf{Ref}(a,\mathsf{cid})=1$ exactly when $a$ authenticates
a reference to $\mathsf{cid}\ne\bot$. Let
$\mathcal{T}_r^{B\rightarrow P}$ contain the broker--provider handoffs
in $\mathcal{T}_r$ whose bilateral signatures and bindings to the
request context, call commitment, and predecessor records verify.
The observable identifier sets are
\begin{equation}
	\begin{aligned}
		\mathcal{Q}_r^{\mathsf{obs}}
		&\triangleq
		\bigl\{\mathsf{cid}\mid
		\exists a\in\mathcal{T}_r\colon
		\mathsf{Ref}(a,\mathsf{cid})=1\bigr\},\\
		\mathcal{Q}_r^{P,\mathsf{obs}}
		&\triangleq
		\bigl\{\mathsf{cid}(a)\mid
		a\in\mathcal{T}_r^{B\rightarrow P}\bigr\}.
	\end{aligned}
	\label{eq:observable-call-set}
\end{equation}
Thus $\mathcal{Q}_r^{P,\mathsf{obs}}\subseteq
\mathcal{Q}_r^{\mathsf{obs}}$.
$\mathcal{R}_{\mathsf{prov}}$ holds exactly when
$\mathcal{R}_{\mathsf{P}}^i=1$ for each
$\mathsf{cid}_i\in\mathcal{Q}_r^{P,\mathsf{obs}}$
(Section~\ref{sec:provider-assurance}).

\noindent\textbf{Receipt coverage and charges.}
Let $N_r(c)$ count settled receipt entries with call identifier $c$.
Exact coverage requires
\[
\forall\,c\colon\quad
N_r(c)=\mathbf{1}[c\in\mathcal{Q}_r^{P,\mathsf{obs}}].
\]
Coverage includes independently submitted acceptances and calls
that fail or abort after acceptance. Each receipt must verify against
its authenticated acceptance, and provider assurance binds
$(\phi_i,\mathbf{u}_i,b_i)$ to that call's execution.
The auditor applies $\Psi_r$ to the verified termination codes to
derive the billable set
$\mathcal{Q}_r^{\mathsf{bill}}\subseteq\mathcal{Q}_r^{P,\mathsf{obs}}$.
Let $\mathcal{F}_r$ be the multiset of auxiliary fee entries in
the settlement. The auditor authenticates each $f\in\mathcal{F}_r$
and checks its request binding, type, issuer, and pricing rule
against $\Psi_r$. It recomputes $\mathsf{cost}(f)$ and checks the
applicable fee cap. With all amounts in a common currency unit,
the recomputed request charge is
\begin{equation}
	B_r^{\star}\triangleq
	\sum_{\mathsf{cid}_i\in\mathcal{Q}_r^{\mathsf{bill}}} b_i
	+\sum_{f\in\mathcal{F}_r}\mathsf{cost}(f).
	\label{eq:request-bill}
\end{equation}
Audit acceptance requires $B_r^U=B_r=B_r^{\star}$, where $B_r^U$
is the charge retained by the user and $B_r$ the settled charge.

\noindent\textbf{Policy compliance.}
Write $\vartheta_i\preceq\Psi_r$ when each call-level allowlist
is a subset of its authorized counterpart and each limit is at most
its bound in $\Psi_r$. Write $(\rho_i,\theta_i)\models\vartheta_i$
when the transformation rule and serving configuration, including
the pricing version, satisfy $\vartheta_i$.
The authenticated routing event fixes $(\rho_i,\theta_i)$.
A verified pre-routing termination sets $(\rho_i,\theta_i)=(\bot,\bot)$.
For provider-accepted calls, $\mathcal{R}_{\mathsf{hand}}$ matches
these values to the opened $B\rightarrow P$ boundary object.
The call-level check is
\begin{equation}
	\mathsf{LocalPolicy}_i\triangleq
	(\vartheta_i\preceq\Psi_r)
	\land
	\bigl(\theta_i\ne\bot\Rightarrow
	(\rho_i,\theta_i)\models\vartheta_i\bigr).
	\label{eq:call-policy-compliance}
\end{equation}
Provider assurance checks deployment binding and assurance-profile
authorization and verifies the transformation in
\eqref{eq:provider-call} on the committed messages.
Each retry uses a fresh call identifier and inherits its
authenticated predecessor's retry-group identifier.
$\mathsf{GlobalPolicy}_r$ checks $|\mathcal{Q}_r^{\mathsf{obs}}|$
and retry counts against $\Psi_r$. It applies token limits to
verified usage for each $\mathsf{cid}_i\in\mathcal{Q}_r^{P,\mathsf{obs}}$
and requires $B_r^\star\leq B_r^{\max}$, where $B_r^{\max}$ is
the signed budget. The authorization relation is
\begin{equation}
	\begin{aligned}
		\mathcal{R}_{\mathsf{auth}}
		&\triangleq
		\mathsf{VerifyAuth}(\mathsf{Auth}_r)\land\mathsf{GlobalPolicy}_r\\
		&\quad{}\land
		\bigl(\forall\,\mathsf{cid}_i\in\mathcal{Q}_r^{\mathsf{obs}}
		\colon \mathsf{LocalPolicy}_i=1\bigr).
	\end{aligned}
	\label{eq:global-authorization}
\end{equation}

\noindent\textbf{Audit semantics.}
With public parameters, the finalized settlement, and user-held
references implicit, the joint validity relation on
$(\mathsf{Auth}_r,\mathcal{T}_r;\mathcal{W})$ is
\begin{equation}
	\mathcal{R}_{\mathsf{TP}}\triangleq
	\mathcal{R}_{\mathsf{auth}}\land\mathcal{R}_{\mathsf{life}}
	\land\mathcal{R}_{\mathsf{hand}}\land\mathcal{R}_{\mathsf{prov}}
	\land\mathcal{R}_{\mathsf{set}}.
	\label{eq:tokenprov-validity}
\end{equation}
The candidate witness $\mathcal{W}$ contains audit payloads,
openings, and a call witness for each observable provider
acceptance. Sections~\ref{sec:authenticated-lifecycle}
and~\ref{sec:closure-settlement} define the lifecycle, handoff,
and settlement checks. The verifier
$\mathsf{Audit}(\mathsf{Auth}_r,\mathcal{T}_r;\mathcal{W}_r)$
uses receipt-bound artifacts for provider assurance and returns
$\delta_r\in\{0,1\}$, with $1$ denoting acceptance.

Index the service positions by
$(E_1,E_2,E_3,E_4)=(U,H,B,P)$. For a reproducible integrity
failure, let $\mathcal{C}_r\subseteq[4]$ index the possible origins
of the violation identified by the first failed check, given the
retained authenticated evidence. The reported responsibility
interval is $\mathsf{Resp}_r=[E_\ell,E_u]$ with
$1\leq\ell\leq u\leq4$.

\subsection{Adversaries and Trust Assumptions}
\label{sec:threat-model}
\noindent\textbf{Adversary and leakage.}
Before request $r$ begins, a probabilistic polynomial-time (PPT)
adversary $\mathcal{A}$ selects which of $H$, $B$, and the providers
to corrupt. It obtains the state and keys they control and
coordinates their actions adaptively during execution and audit.
The honest user $U$ retains its signed authorization and the
delivered response and bill. Write
$\mathsf{View}_{\mathcal{A}}^{\mathsf{real}}$ for the adversary's
audit view. The permitted leakage $\mathsf{Leak}_r$ comprises
its pre-audit state and random tape, together with the ordered
sequence of public records, permitted metadata, authorized
disclosures, and released outputs.

\noindent\textbf{Trust assumptions.}
We rely on $L$ for verifiable inclusion, finality, and consistent
ordering. We assume confidentiality of private evidence in $V$
and the security properties stated in
Section~\ref{sec:preliminaries}. Full provider soundness requires
a user-approved profile that enforces $\mathcal{R}_{\mathsf{P}}^i$,
including integrity checks for covered computation outside an
attested boundary (Section~\ref{sec:provider-assurance}).

\noindent\textbf{Scope.}
\textsc{TokenProv} verifies policy compliance and billing integrity
relative to the authenticated audit transcript $\mathcal{T}_r$.
A policy-compliant and correctly billed workflow may contain
redundant calls.

\subsection{Security Goals}
\label{sec:security-goals}
\noindent\textbf{End-to-end audit soundness.}
For every PPT adversary in the model of
Section~\ref{sec:threat-model}, the audit must satisfy
\begin{equation}
	\Pr\!\left[
	\begin{aligned}
		&\delta_r=1,\\
		&\nexists\,\mathcal{W}\colon
		\mathcal{R}_{\mathsf{TP}}
		(\mathsf{Auth}_r,\mathcal{T}_r;\mathcal{W})=1
	\end{aligned}
	\right]\leq\mathsf{negl}(\lambda).
	\label{eq:audit-soundness}
\end{equation}

\noindent\textbf{Evidence-bounded localization.}
For a reproducible integrity failure with
$\mathcal{C}_r\ne\varnothing$, the reported responsibility interval
$\mathsf{Resp}_r$ must satisfy
\begin{equation}
	\mathcal{C}_r\subseteq\{\ell,\ldots,u\}.
	\label{eq:localization-coverage}
\end{equation}

\noindent\textbf{Audit confidentiality.}
For every PPT adversary $\mathcal{A}$, there exists a PPT
simulator $\mathsf{Sim}$ such that every PPT distinguisher
$\mathcal{D}$ satisfies
\begin{equation}
	\left|
	\begin{aligned}
		&\Pr[\mathcal{D}(\mathsf{View}_{\mathcal{A}}^{\mathsf{real}})=1]\\
		&{}-\Pr[\mathcal{D}(\mathsf{Sim}(1^\lambda,\mathsf{Leak}_r))=1]
	\end{aligned}
	\right|\leq\mathsf{negl}(\lambda).
	\label{eq:audit-confidentiality}
\end{equation}

\section{\textsc{TokenProv} Design}
\label{sec:design}
\begin{figure*}[t]
	\centering
	\includegraphics[width=\linewidth]{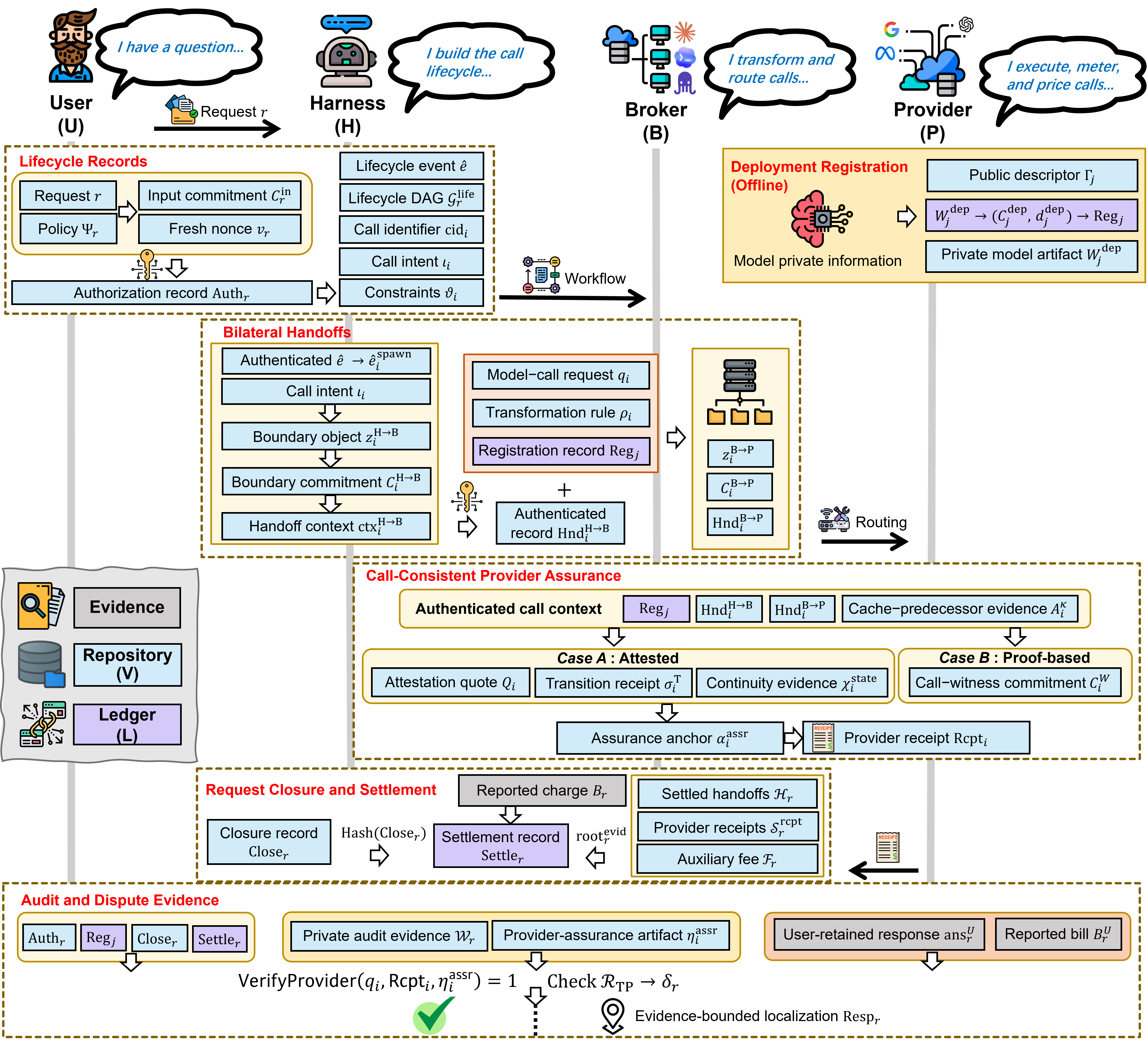}
	\caption{\textsc{TokenProv} workflow from authorization to audit.
		Gray, blue, and purple boxes denote evidence artifacts, off-chain
		records, and ledger anchors, respectively.}
	\label{fig:tokenprov-workflow}
\end{figure*}

Figure~\ref{fig:tokenprov-workflow} shows how \textsc{TokenProv}
authenticates call histories, anchors settlements on $L$, and
uses committed and independently submitted evidence to verify
\eqref{eq:tokenprov-validity}.

\subsection{Lifecycle Records and Handoffs}
\label{sec:authenticated-lifecycle}

\noindent\textbf{Lifecycle records.}
The request lifecycle is an authenticated DAG
$\mathcal{G}_r^{\mathsf{life}}\triangleq
(\mathcal{V}_r^{\mathsf{life}},\mathcal{D}_r^{\mathsf{life}})$
~\cite{moreau2013provdm,bates2015lpm,datta2022alastor}.
Each vertex is a signed event
\begin{equation}
	\begin{aligned}
		e &\triangleq
		(r,h_r^{\mathsf{auth}},\mathsf{eid},\mathsf{cid},
		\mathsf{type},\mathsf{pred},C^{\mathsf{evt}},\mathsf{st}),\\
		\widehat e &\triangleq (e,\mathsf{Sig}_E(e)).
	\end{aligned}
	\label{eq:lifecycle-event}
\end{equation}
Here $E\in\{H,B,P\}$ issues event $\mathsf{eid}$ with type
$\mathsf{type}$, state $\mathsf{st}$, and payload commitment
$C^{\mathsf{evt}}$. An edge $(\widehat e,\widehat e')$ exists
exactly when $\mathsf{Hash}(\widehat e)\in\mathsf{pred}(\widehat e')$.
Request-level events, including response-cache reuse, use
$\mathsf{cid}=\bot$.
For a closed request, $\mathcal{R}_{\mathsf{life}}$ requires valid
event signatures and unique event identifiers. Predecessor references
must resolve within $(r,h_r^{\mathsf{auth}})$. The graph must be
acyclic, with every vertex reachable from a unique root bound to
$\mathsf{Auth}_r$.
Each observable attempt has one $H$-signed spawn event binding
$\iota_i$ from \eqref{eq:call-intent}. Each routed attempt has one
$B$-signed routing event that references its spawn and fixes
$(\rho_i,\theta_i)$. For each observable attempt,
\begin{equation}
	\sum_{\substack{\widehat e\in\mathcal{V}_r^{\mathsf{life}}\\
			\mathsf{cid}(\widehat e)=\mathsf{cid}_i}}
	\mathbf{1}\bigl[\mathsf{st}(\widehat e)\in
	\{\mathsf{succeeded},\mathsf{failed},\mathsf{aborted}\}\bigr]=1.
	\label{eq:unique-terminal-event}
\end{equation}
Retries, rerouting, and failover use fresh call identifiers.
Their event payloads bind the triggering-event reference and the
predecessor's retry-group identifier.
		
\noindent\textbf{Bilateral handoffs.}
For $(X,Y)\in\{(H,B),(B,P)\}$, the sender $X$ commits to the
boundary object $z_i^{X\rightarrow Y}$ as
$(C_i^{X\rightarrow Y},d_i^{X\rightarrow Y})
\leftarrow\mathsf{Commit}(pp,z_i^{X\rightarrow Y})$.
The signed context is
\[
\mathsf{ctx}_i^{X\rightarrow Y}\triangleq
\bigl(\mathsf{DST}_{\mathsf{hnd}},r,h_r^{\mathsf{auth}},
\mathsf{cid}_i,X,Y,C_i^{X\rightarrow Y}\bigr).
\]
The receiver countersigns the context and the sender's signature.
\begin{equation}
	\begin{aligned}
		\sigma_{i,X}^{\mathsf{hnd}}
		&\leftarrow
		\mathsf{Sig}_{X}(\mathsf{ctx}_i^{X\rightarrow Y}),\\
		\sigma_{i,Y}^{\mathsf{hnd}}
		&\leftarrow
		\mathsf{Sig}_{Y}(\mathsf{ctx}_i^{X\rightarrow Y},
		\sigma_{i,X}^{\mathsf{hnd}}),\\
		\mathsf{Hnd}_i^{X\rightarrow Y}
		&\triangleq
		\bigl(\mathsf{ctx}_i^{X\rightarrow Y},
		\sigma_{i,X}^{\mathsf{hnd}},\sigma_{i,Y}^{\mathsf{hnd}}\bigr).
	\end{aligned}
	\label{eq:bilateral-handoff}
\end{equation}
Let $\widehat e_i^{\mathsf{spawn}}$ be the unique spawn event for
attempt $i$ and $\mathsf{Reg}_j$ the deployment registration for
$\theta_i$ (Section~\ref{sec:provider-assurance}).
The two boundary objects are
\begin{equation}
	\begin{aligned}
		z_i^{H\rightarrow B}
		&\triangleq
		\bigl(\iota_i,\mathsf{Hash}(\widehat e_i^{\mathsf{spawn}})\bigr),\\
		z_i^{B\rightarrow P}
		&\triangleq
		\bigl(q_i,\rho_i,\mathsf{Hash}(\mathsf{Hnd}_i^{H\rightarrow B}),
		\mathsf{Hash}(\mathsf{Reg}_j)\bigr).
	\end{aligned}
	\label{eq:call-boundary-objects}
\end{equation}
Before countersigning, $Y$ verifies the opening, sender signature,
context, and predecessor bindings. The provider also checks
finalized inclusion of $\mathsf{Reg}_j$ on $L$. Retransmissions
reuse the existing acceptance. For each
$(r,h_r^{\mathsf{auth}},\mathsf{cid}_i)$, the audit permits at most
one distinct acceptance per handoff direction across all recipients.
For each $\mathsf{cid}_i\in\mathcal{Q}_r^{P,\mathsf{obs}}$,
the authenticated references establish
\begin{equation}
	\mathsf{Auth}_r
	\prec_{\mathsf{prot}}\widehat e_i^{\mathsf{spawn}}
	\prec_{\mathsf{prot}}\mathsf{Hnd}_i^{H\rightarrow B}
	\prec_{\mathsf{prot}}\mathsf{Hnd}_i^{B\rightarrow P}.
	\label{eq:handoff-order}
\end{equation}
$\mathcal{R}_{\mathsf{hand}}$ checks signatures, openings, context
and predecessor bindings, and acceptance uniqueness for all handoffs
in $\mathcal{T}_r$. For each $B\rightarrow P$ handoff, it matches the
opened $(\rho_i,\theta_i)$ to the authenticated routing event.
Both endpoints retain the signed handoff, its boundary object and
opening, and supporting predecessor records for the period in
$\Psi_r$. A holder may submit this evidence directly under the
rules in Section~\ref{sec:authorization-boundary}.

\subsection{Call-Consistent Provider Assurance}
\label{sec:provider-assurance}
For each $\mathsf{cid}_i\in\mathcal{Q}_r^{P,\mathsf{obs}}$,
the provider assurance relation for the authenticated acceptance
$\mathsf{Hnd}_i^{B\rightarrow P}$ is
\begin{equation}
	\mathcal{R}_{\mathsf{P}}^{i}
	\triangleq
	\mathcal{R}_{\mathsf{dep}}^{i}
	\land \mathcal{R}_{\mathsf{in}}^{i}
	\land \mathcal{R}_{\mathsf{exec}}^{i}
	\land \mathcal{R}_{\mathsf{acct}}^{i}
	\land \mathcal{R}_{\mathsf{link}}^{i}.
	\label{eq:provider-composite-relation}
\end{equation}
The linking predicate $\mathcal{R}_{\mathsf{link}}^i$ checks equality
of values across the component relations and binds the verified
result, usage, and charge to the receipt.
Table~\ref{tab:provider-relations} summarizes the checks.

\begin{table}[!t]
	\centering
	\caption{Checks in the call-level provider relation.}
	\label{tab:provider-relations}
	\footnotesize
	\renewcommand{\arraystretch}{1.06}
	\setlength{\tabcolsep}{4pt}
	\begin{tabularx}{\columnwidth}{
			@{}>{\centering\arraybackslash}p{0.18\columnwidth}Y@{}
		}
		\toprule
		\makebox[\linewidth][c]{\textbf{Relation}}
		&
		\makebox[\linewidth][c]{\textbf{Enforced consistency}} \\
		\midrule
		$\mathcal{R}_{\mathsf{dep}}^{i}$
		& Signed policy $\leftrightarrow$ deployment and assurance profile \\
		$\mathcal{R}_{\mathsf{in}}^{i}$
		& Handoffs $\leftrightarrow$ transformation $\leftrightarrow$ model input \\
		$\mathcal{R}_{\mathsf{exec}}^{i}$
		& Input and cache $\leftrightarrow$ execution $\leftrightarrow$ result \\
		$\mathcal{R}_{\mathsf{acct}}^{i}$
		& Execution $\leftrightarrow$ metered usage $\leftrightarrow$ charge \\
		$\mathcal{R}_{\mathsf{link}}^{i}$
		& Shared-value equality and receipt binding \\
		\bottomrule
	\end{tabularx}
\end{table}

\noindent\textbf{Deployment registration.}
For deployment $j$, the provider publishes the descriptor
\[
\Gamma_j\triangleq
\bigl(\mathsf{mid}_j,\mathsf{ver}_j,
h_j^{\mathsf{model}},h_j^{\mathsf{tok}},
h_j^{\mathsf{canon}},h_j^{\mathsf{serv}},\gamma_j\bigr).
\]
The four digests bind the model artifacts, tokenizer, canonicalizer,
and serving specification, respectively. The serving specification
fixes the decoding rules and the versioned cache, metering, and pricing
rules selected by $\theta_i$ (Section~\ref{sec:serving-semantics}).
These rules also specify cache-object identifiers and the treatment
of returned reasoning tokens. The assurance profile $\gamma_j$
identifies the relation in \eqref{eq:provider-composite-relation}
and its version. It fixes the public-parameter and verification-key
digests for proving, or the program measurement and protected boundary
for attestation.

Let $W_j^{\mathsf{dep}}$ contain the deployment artifacts and
configuration. Under the commitment parameters $pp_{\mathsf{P}}$
fixed by $\gamma_j$, the provider computes
$(C_j^{\mathsf{dep}},d_j^{\mathsf{dep}})
\leftarrow\mathsf{Commit}(pp_{\mathsf{P}},W_j^{\mathsf{dep}})$
and issues
\begin{equation}
	\begin{aligned}
		\mathsf{reg}_j &\triangleq
		\bigl(\mathsf{pid}_j,\Gamma_j,C_j^{\mathsf{dep}}\bigr),\\
		\mathsf{Reg}_j &\triangleq
		\bigl(\mathsf{reg}_j,\mathsf{Sig}_{P}(\mathsf{reg}_j)\bigr).
	\end{aligned}
	\label{eq:deployment-registration}
\end{equation}
The $B\rightarrow P$ handoff references $\mathsf{Hash}(\mathsf{Reg}_j)$
after registration is finalized on $L$.
$\mathcal{R}_{\mathsf{dep}}^i$ checks that $\theta_i$ matches the
registration and that $\Psi_r$ authorizes $h_j^{\mathsf{model}}$
and $\gamma_j$. A provenance credential used for authorization
must bind $h_j^{\mathsf{model}}$ under an issuer approved by $\Psi_r$.
The assurance backend verifies the opening of $C_j^{\mathsf{dep}}$
and recomputes the four digests from the deployment witness used by
$\mathcal{R}_{\mathsf{exec}}^i$.

\noindent\textbf{Call execution and accounting.}
$\mathcal{R}_{\mathsf{in}}^i$ checks both handoff openings
against \eqref{eq:call-boundary-objects}. It verifies the
transformation in \eqref{eq:provider-call} on the opened messages
and checks \eqref{eq:provider-input} with the registered
canonicalizer and tokenizer.

The cache evidence $A_i^\kappa$ binds $\kappa_i^{-}$ to an
admissible initial state or a verified predecessor post-state
for the same cache object, deployment, and cache-rule version.
$\mathcal{R}_{\mathsf{exec}}^i$ verifies this evidence and checks
\eqref{eq:provider-execution} and \eqref{eq:logical-cache-step}
using the committed deployment witness. Immutable prefix entries
permit concurrent reads. Conflicting updates must satisfy the
registered state-continuity rules.

$\mathcal{R}_{\mathsf{acct}}^i$ checks $\mathbf{u}_i$ and $b_i$
under the registered metering and pricing rules in
Section~\ref{sec:serving-semantics}. $\mathcal{R}_{\mathsf{link}}^i$
matches the verified input and configuration to the execution.
It binds accounting to that execution's trace $g_i$, returned
tokens $\boldsymbol{\tau}_i^{\mathsf{out}}$, and cache states
$(\kappa_i^{-},\kappa_i^{+})$.

\noindent\textbf{Provider receipts.}
For each accepted call, the assurance mode
$\mathsf{mode}_i\in\{\mathsf{zk},\mathsf{tee}\}$ must satisfy
$\Psi_r$ and match the registered profile $\gamma_j$.
The provider commits to the execution result as
\[
(C_i^{\mathsf{res}},d_i^{\mathsf{res}})
\leftarrow
\mathsf{Commit}\bigl(
pp_{\mathsf{P}},
(\boldsymbol{\tau}_i^{\mathsf{out}},\phi_i,\kappa_i^{+})
\bigr).
\]
In proof-based mode, the call witness $W_i^{\mathsf{call}}$
contains the values and openings required by $\mathcal{R}_{\mathsf{P}}^i$,
including $A_i^\kappa$, $g_i$, $\xi_i$, and $d_i^{\mathsf{res}}$.
The provider computes
\[
(C_i^W,d_i^W)
\leftarrow\mathsf{Commit}(pp_{\mathsf{P}},W_i^{\mathsf{call}}).
\]

The deterministic call nullifier~\cite{bensasson2014zerocash} is
\begin{equation}
	\mathsf{nf}_i\triangleq
	\mathsf{Hash}\bigl(\mathsf{DST}_{\mathsf{call}},
	h_r^{\mathsf{auth}},\mathsf{cid}_i\bigr).
	\label{eq:provider-call-nullifier}
\end{equation}
Let $\eta_i^{\mathsf{assr}}$ be the assurance artifact for the
selected mode. The receipt binds the anchor
\begin{equation}
	\alpha_i^{\mathsf{assr}}\triangleq
	\begin{cases}
		C_i^W, & \mathsf{mode}_i=\mathsf{zk},\\
		\mathsf{Hash}(\eta_i^{\mathsf{assr}}),
		& \mathsf{mode}_i=\mathsf{tee}.
	\end{cases}
	\label{eq:provider-assurance-anchor}
\end{equation}
The receipt body and signed receipt are
\begin{equation}
	\begin{aligned}
		\mathsf{rcpt}_i
		&\triangleq
		\bigl(r,\mathsf{cid}_i,\mathsf{nf}_i,
		\mathsf{Hash}(\mathsf{Reg}_j),C_i^{B\rightarrow P},\\
		&\qquad
		\mathsf{Hash}(A_i^\kappa),C_i^{\mathsf{res}},
		\phi_i,\mathbf{u}_i,b_i,\mathsf{mode}_i,
		\alpha_i^{\mathsf{assr}}\bigr),\\
		\mathsf{Rcpt}_i
		&\triangleq
		\bigl(\mathsf{rcpt}_i,\mathsf{Sig}_P(\mathsf{rcpt}_i)\bigr).
	\end{aligned}
	\label{eq:provider-receipt}
\end{equation}
The auditor verifies the receipt signature under the provider key
authenticated by $\mathsf{Hnd}_i^{B\rightarrow P}$ and recomputes
$\mathsf{nf}_i$. $\mathcal{R}_{\mathsf{link}}^i$ checks the deployment,
handoff, and cache-predecessor bindings. It verifies that
$C_i^{\mathsf{res}}$ opens to the verified result and that
$(\phi_i,\mathbf{u}_i,b_i)$ match that execution.
Both $P$ and $B$ retain the signed receipt for settlement.
In proof-based mode, the finalized receipt fixes $C_i^W$ before
the audit challenge.

Both assurance realizations establish the same call-level relation
$\mathcal{R}_{\mathsf{P}}^i$.

\begin{assurancecase}
	\noindent\textbf{\textit{Case A.} Attested assurance (TEE-enabled deployments).}
	The protected transition in Section~\ref{sec:attested-execution}
	uses $D=\Gamma_j$ and the program measurement specified by $\gamma_j$.
	Its authenticated input context and output are
	\begin{equation}
		\begin{aligned}
			x_i^T &\triangleq
			\bigl(\Psi_r,\mathsf{Reg}_j,
			\mathsf{Hnd}_i^{H\rightarrow B},
			\mathsf{Hnd}_i^{B\rightarrow P},A_i^\kappa\bigr),\\
			y_i^T &\triangleq
			\bigl(C_i^{\mathsf{res}},\phi_i,
			\mathbf{u}_i,b_i,\mathsf{nf}_i\bigr).
		\end{aligned}
		\label{eq:provider-attested-io}
	\end{equation}
	The measured program enforces $\mathcal{R}_{\mathsf{P}}^i$.
	For covered computation outside the protected boundary, it verifies
	an integrity proof $\pi_i^{\mathsf{ext}}$ using the verifier
	specified by $\gamma_j$ and checks its binding to the same call
	inputs and outputs. If all covered computation runs inside the
	boundary, $\pi_i^{\mathsf{ext}}=\bot$.
	
	Let $Q_i$ be the quote, $\sigma_i^T$ the transition signature,
	and $\chi_i^{\mathsf{state}}$ the state digests and continuity
	evidence. The assurance artifact is
	\[
	\eta_i^{\mathsf{assr}}\triangleq
	\bigl(Q_i,\sigma_i^T,\chi_i^{\mathsf{state}},
	\pi_i^{\mathsf{ext}}\bigr).
	\]
	The auditor validates $Q_i$ and checks the transition on
	$(x_i^T,y_i^T)$ using \eqref{eq:attested-transition-check}.
	It verifies any required external proof and checks
	$\mathsf{Hash}(\eta_i^{\mathsf{assr}})=\alpha_i^{\mathsf{assr}}$.
\end{assurancecase}

\begin{assurancecase}
	\noindent\textbf{\textit{Case B.} Proof-based assurance (general deployments).}
	The provider instantiates the commit-and-prove argument of
	Section~\ref{sec:commit-and-prove} for $\mathcal{R}_{\mathsf{P}}^i$
	with public statement
	\[
	x_i^{\mathsf{zk}}\triangleq
	\bigl(\Psi_r,\mathsf{Reg}_j,
	\mathsf{Hnd}_i^{H\rightarrow B},
	\mathsf{Hnd}_i^{B\rightarrow P},
	\mathsf{Rcpt}_i,\mathsf{Hash}(A_i^\kappa)\bigr).
	\]
	The provider generates
	\begin{equation}
		\pi_i^{\mathsf{zk}}
		\leftarrow
		\mathsf{Prove}\bigl(
		pp_{\mathsf{P}},x_i^{\mathsf{zk}},C_i^W;
		W_i^{\mathsf{call}},d_i^W\bigr).
		\label{eq:provider-zk-proof}
	\end{equation}
	The assurance artifact is
	$\eta_i^{\mathsf{assr}}\triangleq(C_i^W,\pi_i^{\mathsf{zk}})$.
	Using the authenticated statement and the verification parameters
	fixed by $\gamma_j$, the auditor requires
	\[
	C_i^W=\alpha_i^{\mathsf{assr}},
	\qquad
	\mathsf{Verify}\bigl(
	pp_{\mathsf{P}},x_i^{\mathsf{zk}},C_i^W,
	\pi_i^{\mathsf{zk}}\bigr)=1.
	\]
\end{assurancecase}

For either realization, the auditor uses $\gamma_j$ to select the
verifier for $\mathsf{mode}_i$ and checks its registered parameter
digests. With the authenticated call context and verification
parameters implicit, provider assurance requires
\begin{equation}
	\mathsf{VerifyProvider}\bigl(
	q_i,\mathsf{Rcpt}_i,\eta_i^{\mathsf{assr}}\bigr)=1.
	\label{eq:unified-provider-verification}
\end{equation}
For deferred proving, $P$ retains $W_i^{\mathsf{call}}$ and
$d_i^W$ for the retention period in $\Psi_r$.

\subsection{Request Closure and Settlement}
\label{sec:closure-settlement}
\noindent\textbf{Request closure.}
Once all recorded attempts have terminated, $H$ commits to the
final response $\mathsf{ans}_r$:
\[
(C_r^{\mathsf{ans}},d_r^{\mathsf{ans}})
\leftarrow \mathsf{Commit}(pp,\mathsf{ans}_r).
\]
$H$ appends a unique signed finalization event with $\mathsf{cid}=\bot$.
The event binds $C_r^{\mathsf{ans}}$ and is causally downstream of
every call-terminal event. The lifecycle root is
\[
\mathsf{root}_r^{\mathsf{life}}\triangleq
\mathsf{MRoot}\bigl(\mathsf{Sort}(\mathcal{V}_r^{\mathsf{life}})\bigr),
\]
where $\mathsf{Sort}$ orders events by their canonically encoded
identifiers. The Merkle construction binds the sequence length
and leaf positions~\cite{merkle1988digital,laurie2021ct}.
$H$ then issues the signed closure record
\begin{equation}
	\begin{aligned}
		\mathsf{close}_r &\triangleq
		(r,h_r^{\mathsf{auth}},\mathsf{root}_r^{\mathsf{life}},
		C_r^{\mathsf{ans}}),\\
		\mathsf{Close}_r &\triangleq
		(\mathsf{close}_r,\mathsf{Sig}_H(\mathsf{close}_r)).
	\end{aligned}
	\label{eq:request-closure}
\end{equation}
$\mathcal{R}_{\mathsf{close}}$ checks the signature, request context,
and finalization conditions, and recomputes the lifecycle root
from the complete event sequence. It requires
$\mathsf{Open}(pp,C_r^{\mathsf{ans}},\mathsf{ans}_r^U,d_r^{\mathsf{ans}})=1$
for the response $\mathsf{ans}_r^U$ retained by the user.

\noindent\textbf{Settlement evidence.}
Let $\mathcal{H}_r$ and $\mathcal{S}_r^{\mathsf{rcpt}}$ denote
the multisets of handoff records and provider receipts submitted
for settlement. The settlement evidence root is
\begin{equation}
	\mathsf{root}_r^{\mathsf{evid}}
	\triangleq
	\mathsf{MRoot}\bigl(\mathsf{Sort}\bigl(\mathsf{Tag}(
	\mathcal{H}_r \uplus \mathcal{S}_r^{\mathsf{rcpt}}
	\uplus \mathcal{F}_r)\bigr)\bigr).
	\label{eq:settlement-evidence-root}
\end{equation}
Here $\uplus$ denotes multiset union. $\mathsf{Tag}$ labels each
entry by record type, and $\mathsf{Sort}$ orders the tagged
entries by canonical encoding. The auditor recomputes the root
from all committed entries, preserving their multiplicities.

\noindent\textbf{Settlement record.}
The authorization digest determines the settlement nullifier
\[
\mathsf{nf}_r^{\mathsf{set}}
\triangleq
\mathsf{Hash}\bigl(
\mathsf{DST}_{\mathsf{set}},h_r^{\mathsf{auth}}\bigr).
\]
The broker and harness authenticate the settlement record
\begin{equation}
	\begin{aligned}
		\mathsf{settle}_r
		&\triangleq
		\bigl(r,\mathsf{nf}_r^{\mathsf{set}},h_r^{\mathsf{auth}},
		\mathsf{Hash}(\mathsf{Close}_r),
		\mathsf{root}_r^{\mathsf{evid}},B_r\bigr),\\
		\mathsf{Settle}_r
		&\triangleq
		\bigl(\mathsf{settle}_r,\sigma_r^{B},
		\mathsf{Sig}_{H}(\mathsf{settle}_r,\sigma_r^{B})\bigr),
	\end{aligned}
	\label{eq:settlement-record}
\end{equation}
where $\sigma_r^{B}\leftarrow\mathsf{Sig}_{B}(\mathsf{settle}_r)$.
The ledger $L$ atomically rejects a second initial settlement
with the same nullifier.

\noindent\textbf{Settlement validity.}
With the audit inputs implicit, define
\begin{equation}
	\begin{aligned}
		\mathcal{R}_{\mathsf{set}}
		\triangleq
		\mathcal{R}_{\mathsf{close}}
		\land\mathcal{R}_{\mathsf{cover}}
		\land\mathcal{R}_{\mathsf{charge}}
		\land\mathcal{R}_{\mathsf{uniq}}
		\land
		[B_r^{U}=B_r=B_r^{\star}].
	\end{aligned}
	\label{eq:settlement-relation}
\end{equation}
The predicates $\mathcal{R}_{\mathsf{cover}}$ and
$\mathcal{R}_{\mathsf{charge}}$ enforce the receipt coverage
and billing rules in Section~\ref{sec:authorization-boundary},
respectively. The predicate $\mathcal{R}_{\mathsf{uniq}}$ rejects
duplicate call identifiers or nullifiers among settled receipts,
duplicate auxiliary-fee claims, and repeated initial settlements.

\subsection{Audit and Dispute Evidence}
\label{sec:end-to-end-audit}
\noindent\textbf{Audit procedure.}
After $\mathsf{Settle}_r$ is finalized on $L$, an authorized
auditor selects $r$ for audit with probability $p_r$ using
fresh randomness. A user challenge within the audit period
specified by $\Psi_r$ triggers the procedure.
The auditor reconstructs $\mathcal{T}_r$ as specified in
Section~\ref{sec:authorization-boundary}.
Algorithm~\ref{alg:end-to-end-audit} verifies finalized
ledger inclusion and checks $\mathcal{R}_{\mathsf{TP}}$ using
$\mathcal{T}_r$, $\mathcal{W}_r$, and the user-retained records
$(\mathsf{Auth}_r,\mathsf{ans}_r^{U},B_r^{U})$.
The audit returns $\delta_r=1$ exactly when all checks pass
and the required evidence is supplied by the response deadline
in $\Psi_r$.

\begin{algorithm}[t]
	\caption{Request-level audit for $r$.}
	\label{alg:end-to-end-audit}
	\begin{algorithmic}[1]
		\fontsize{9.5pt}{11pt}\selectfont
		\Require $\mathsf{Auth}_r$, $\mathsf{Settle}_r$,
		$\mathcal{T}_r$, $\mathcal{W}_r$, $\mathsf{ans}_r^{U}$, $B_r^{U}$
		\Ensure $\delta_r\in\{0,1\}$
		
		\ShadedComment{\Comment{\textit{Authentication and commitments}}}
		\State \textbf{assert} $\mathsf{VerifyAuth}(\mathsf{Auth}_r)=1$
		\State \textbf{assert} $\mathsf{Settle}_r\in L$
		\Comment{Finalized inclusion}
		\State \textbf{assert} $\mathsf{VerifySig}_{pk_B}
		(\mathsf{Hash}(\mathsf{settle}_r),\sigma_r^{B})=1$
		\State \textbf{assert} $\mathsf{VerifySig}_{pk_H}
		(\mathsf{Hash}(\mathsf{settle}_r,\sigma_r^{B}),
		\mathsf{Settle}_r[3])=1$
		\State Recompute $h_r^{\mathsf{auth}}$ and $\mathsf{root}_r^{\mathsf{evid}}$
		\Comment{\eqref{eq:authorization-digest}, \eqref{eq:settlement-evidence-root}}
		\State $\mathsf{nf}_r^{\mathsf{set}}\gets
		\mathsf{Hash}(\mathsf{DST}_{\mathsf{set}},h_r^{\mathsf{auth}})$
		\State Check $\mathsf{settle}_r$ against \eqref{eq:settlement-record}
		
		\ShadedComment{\Comment{\textit{Call history and receipt coverage}}}
		\State Derive $\mathcal{Q}_r^{\mathsf{obs}}$ and
		$\mathcal{Q}_r^{P,\mathsf{obs}}$ \Comment{\eqref{eq:observable-call-set}}
		\State \textbf{assert}
		$\mathcal{R}_{\mathsf{life}}\land\mathcal{R}_{\mathsf{hand}}$
		\State \textbf{assert}
		$\forall\,\mathsf{cid}_i\in\mathcal{Q}_r^{\mathsf{obs}}\colon
		\mathsf{LocalPolicy}_i=1$
		\State \textbf{assert}
		$\mathcal{R}_{\mathsf{close}}\land\mathcal{R}_{\mathsf{cover}}
		\land\mathcal{R}_{\mathsf{uniq}}$
		
		\ShadedComment{\Comment{\textit{Provider assurance and bill verification}}}
		\ForAll{$\mathsf{cid}_i\in\mathcal{Q}_r^{P,\mathsf{obs}}$}
		\State Receive $\eta_i^{\mathsf{assr}}$ from $P$
		\Comment{Deadline in $\Psi_r$}
		\State \textbf{assert} $\eta_i^{\mathsf{assr}}\ne\bot$
		\State \textbf{assert} $\mathsf{VerifyProvider}
		(q_i,\mathsf{Rcpt}_i,\eta_i^{\mathsf{assr}})=1$
		\EndFor
		\State \textbf{assert} $\mathcal{R}_{\mathsf{charge}}$
		\State Compute $B_r^\star$ \Comment{\eqref{eq:request-bill}}
		\State \textbf{assert} $\mathsf{GlobalPolicy}_r=1$
		\State \textbf{assert} $B_r^{U}=B_r=B_r^\star$
		\State \Return $1$
	\end{algorithmic}
\end{algorithm}

\noindent\textbf{Proving cost under sampling.}
For a fixed workload with at most one new proof per call in a
sampled audit, the expected proving time per finalized
request due to sampling is
\[
p_r \sum_{\mathsf{cid}_i\in\mathcal{Q}_r^{P,\mathsf{obs}}}
t_i^{\mathsf{prove}},
\]
where $t_i^{\mathsf{prove}}$ is the proof-generation time for
call $i$ and is zero if no new proof is generated.

\noindent\textbf{Dispute evidence.}
On $\delta_r=0$, the auditor records the first failed check in
\begin{equation}
	F_r\triangleq
	\bigl(r,\mathsf{Hash}(\mathsf{Settle}_r),
	\mathsf{check},\mathsf{reason},\mathsf{refs}\bigr).
	\label{eq:audit-failure-bundle}
\end{equation}
An integrity failure is reproducible if an authorized reviewer
obtains the same rejection by rerunning $\mathsf{check}$ on the
authenticated inputs identified by $\mathsf{refs}$.
Missing required evidence at the response deadline yields an
incomplete audit. If the challenge and its deadline are anchored
on $L$, the reviewer uses ledger order to check whether the
response commitment was posted on time.

For a reproducible integrity failure, the auditor reports a
responsibility interval $\mathsf{Resp}_r$ satisfying
\eqref{eq:localization-coverage}. Two distinct $H$-signed
spawn events for the same
$(r,h_r^{\mathsf{auth}},\mathsf{cid}_i)$ violate spawn uniqueness
and yield $\mathsf{Resp}_r=[H,H]$. If every role remains a
candidate, the auditor reports $[U,P]$. Corrected invoices and
dispute outcomes are appended as separate records referencing
$\mathsf{Hash}(\mathsf{Settle}_r)$.

\section{Security Analysis}
\label{sec:security}

We analyze Algorithm~\ref{alg:end-to-end-audit} under the
assumptions of Section~\ref{sec:threat-model}. The integrity
guarantees apply to the authenticated transcript $\mathcal{T}_r$.
Appendix~\ref{app:security-proofs} gives the proofs.

\subsection{Integrity and Billing Guarantees}
\label{sec:integrity-security}

\begin{lemma}[Authenticated structural integrity]
	\label{lem:structural-integrity}
	Except with negligible probability, an accepting audit
	establishes a valid lifecycle rooted in $\mathsf{Auth}_r$,
	with exactly one spawn and one terminal event for each
	$\mathsf{cid}_i\in\mathcal{Q}_r^{\mathsf{obs}}$.
	Each required handoff binds its opened boundary object and
	predecessor references to the request, call identifier, and
	transfer direction. The finalized Merkle roots bind the complete
	opened lifecycle and settlement record sequences.
\end{lemma}

\begin{lemma}[Provider-assurance soundness]
	\label{lem:provider-soundness}
	For any PPT adversary, consider an adaptively designated call
	$\mathsf{cid}_i\in\mathcal{Q}_r^{P,\mathsf{obs}}$ checked
	under a user-approved profile enforcing
	\eqref{eq:provider-composite-relation}. With the authenticated
	public inputs implicit, the check satisfies
	\begin{equation}
		\Pr\!\left[
		\begin{aligned}
			&\mathsf{VerifyProvider}
			(q_i,\mathsf{Rcpt}_i,\eta_i^{\mathsf{assr}})=1,\\
			&\nexists\,w_i\colon \mathcal{R}_{\mathsf{P}}^{i}(w_i)=1
		\end{aligned}
		\right]\leq\mathsf{negl}(\lambda),
		\label{eq:provider-soundness}
	\end{equation}
	where $w_i$ ranges over call witnesses.
\end{lemma}

\begin{lemma}[Settlement exactness]
	\label{lem:settlement-exactness}
	Except with negligible probability, $\delta_r=1$ implies
	\[
	\bigl(\forall\,c\colon
	N_r(c)=\mathbf{1}[c\in\mathcal{Q}_r^{P,\mathsf{obs}}]\bigr)
	\land [B_r^{U}=B_r=B_r^{\star}].
	\]
\end{lemma}

\begin{theorem}[End-to-end audit soundness]
	\label{thm:audit-soundness}
	Under the assumptions of Section~\ref{sec:threat-model},
	Algorithm~\ref{alg:end-to-end-audit} satisfies
	\eqref{eq:audit-soundness}.
\end{theorem}

\begin{proof}[Proof sketch]
	Lemmas~\ref{lem:structural-integrity}
	and~\ref{lem:provider-soundness} fix the authenticated records
	and supply a witness for each observable provider acceptance.
	Commitment binding and cache-predecessor checks make these
	witnesses consistent with one another and with the openings
	in $\mathcal{W}_r$. The authorization and settlement checks,
	including the exactness conditions of
	Lemma~\ref{lem:settlement-exactness}, then establish
	\eqref{eq:tokenprov-validity} for the same joint witness.
	A union bound over the polynomially many checks yields
	\eqref{eq:audit-soundness}.
\end{proof}

\subsection{Deferred Proving and Localization}
\label{sec:deferred-accountability}
\begin{proposition}[Challenge-independent witness fixation]
	\label{prop:deferred-proof}
	Fix a proof-based receipt whose commitment $C_i^W$ is bound
	by the finalized settlement before audit selection or a user
	challenge. For any PPT prover, the witnesses extracted from
	accepted deferred proofs for that receipt within a single audit
	agree on $W_i^{\mathsf{call}}$, except with negligible probability.
\end{proposition}

\noindent\textbf{Evidence-bounded localization.}
For a reproducible integrity failure with
$\mathcal{C}_r\ne\varnothing$, exclusions justified by authenticated
evidence retain every position in $\mathcal{C}_r$. The interval
spanning the remaining positions therefore satisfies
\eqref{eq:localization-coverage}.

\subsection{Audit Confidentiality}
\label{sec:security-confidentiality}

\begin{theorem}[Audit confidentiality]
	\label{thm:audit-confidentiality}
	Assume confidentiality of private evidence in $V$ and of
	protected execution state. If private audit channels restrict
	plaintext to authorized recipients and expose only metadata
	permitted by $\mathsf{Leak}_r$, the audit satisfies
	\eqref{eq:audit-confidentiality}.
\end{theorem}

The simulator restores the adversary's leaked pre-audit state,
including its random tape, and replays the ordered audit
observations with the exact encodings of disclosed artifacts.
Appendix~\ref{app:proof-confidentiality} shows that this replay
reproduces the adversary's audit view.

\section{Evaluation}
\label{sec:evaluation}
We evaluate the cost and scalability of \textsc{TokenProv}
and test its ability to detect provenance violations.

\subsection{Experimental Setup}
\label{sec:evaluation-setup}
\noindent\textbf{Testbed.}
We co-locate the Python protocol roles on a shared server with
two Intel Xeon 6747P CPUs and eight NVIDIA B300 GPUs.
Each model service uses one GPU and exposes an OpenAI-compatible
HTTP API. We use Transformers~4.54.0 and PyTorch~2.12.0+cu130
with scaled dot-product attention. The prototype uses Ed25519
signatures~\cite{bernstein2012highspeed}, authenticated Merkle
structures, and a local authenticated append-only ledger.
Table~\ref{tab:default-config} summarizes the serving settings.
We benchmark isolated-key receipt signing on a separate laptop.

\begin{table}[!t]
	\centering
	\caption{Default experimental settings.}
	\label{tab:default-config}
	\footnotesize
	\setlength{\tabcolsep}{2.5pt}
	\renewcommand{\arraystretch}{1.04}
	
	\hrule height \heavyrulewidth
	\vspace{2pt}
	\noindent
	\begin{minipage}[t]{0.50\columnwidth}
		\begin{tabularx}{\linewidth}[t]{>{\raggedright\arraybackslash}Xr}
			\rowcolor{gray!12}
			\multicolumn{2}{l}{\textbf{Serving configuration}} \\
			Precision            & \texttt{bfloat16} \\
			Context limit        & 4,096 tokens \\
			Input tokens/call    & 2,048 \\
			Max.\ output/call    & 256 tokens \\
			Concurrency          & 8 \\
			Microbatch size      & 8 \\
			Batching window      & 2\,ms \\
			Decoding             & Greedy \\
			\addlinespace[4pt]
			\rowcolor{gray!12}
			\multicolumn{2}{l}{\textbf{Workflow configuration}} \\
			Model calls          & 8 (sequential) \\
			Lifecycle events     & 32 \\
			Retries/failovers    & None \\
		\end{tabularx}
	\end{minipage}\hfill%
	\begin{minipage}[t]{0.48\columnwidth}
		\begin{tabularx}{\linewidth}[t]{>{\raggedright\arraybackslash}Xr}
			\rowcolor{gray!12}
			\multicolumn{2}{l}{\textbf{Broker model pool}} \\
			\multicolumn{2}{l}{Qwen3-1.7B (reference)} \\
			\multicolumn{2}{l}{Qwen3-4B} \\
			\multicolumn{2}{l}{Qwen3-14B} \\
			\multicolumn{2}{l}{Phi-3-mini-4k-instruct (3.8B)} \\
			\multicolumn{2}{>{\raggedright\arraybackslash}%
				p{\dimexpr\linewidth-2\tabcolsep\relax}}%
			{DeepSeek-R1-Distill-\mbox{Llama-8B}} \\
			\addlinespace[2pt]
			\rowcolor{gray!12}
			\multicolumn{2}{l}{\textbf{Measurement configuration}} \\
			Measured requests          & 128 \\
			Warm-up requests           & 4 \\
			Rounds per model           & 3 \\
			Scheme-order seed          & 42 \\
			\addlinespace[2pt]
			\rowcolor{gray!12}
			\multicolumn{2}{l}{\textbf{Audit and settlement}} \\
			Auxiliary fees per request & 0 \\
		\end{tabularx}
	\end{minipage}\par
	\vspace{2pt}
	\hrule height \heavyrulewidth
	\vspace{2pt}
	\parbox{\columnwidth}{\scriptsize
		Request counts are per scheme per round.
		Warm-ups are excluded.}
\end{table}

\begin{figure*}[t]
	\centering
	\includegraphics[width=\textwidth]{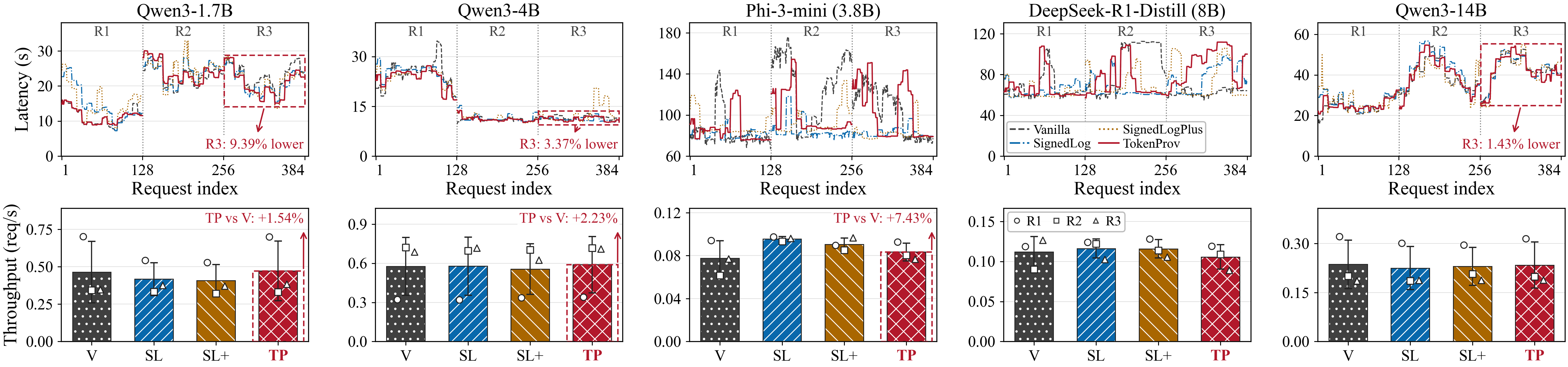}
	\caption{Request latency by submission index (top) and batch
		throughput (bottom) at concurrency eight. Throughput markers
		show individual rounds. Bars and whiskers show the mean and
		sample standard deviation across three rounds, respectively.}
	\label{fig:e2e-latency}
\end{figure*}

\noindent\textbf{Models and workloads.}
We evaluate five models spanning three backbone families and
1.7B--14B parameters. The three Qwen3 variants support
comparisons across model sizes within one family~\cite{yang2025qwen3}.
The pool also includes the instruction-tuned
Phi-3-mini-4k-instruct~\cite{abdin2024phi3} and the reasoning-distilled
DeepSeek-R1-Distill-Llama-8B~\cite{guo2025deepseekr1}.
The serving benchmarks use sequential calls on synthetic prompts
with fixed input lengths after chat templating and tokenization.
We disable cross-call prefix caching in these runs.
Figure~\ref{fig:tokenprovenancecaseboxes} in
Appendix~\ref{app:additional-evaluation} traces how tokens are
generated and metered for three example requests, and how
signed usage records support settlement and offline audit.

\noindent\textbf{Baselines.}
\textsc{Vanilla} runs the same workload without provenance
protection. \textsc{SignedLog} authenticates individual records.
\textsc{SignedLogPlus} uses disclosed records to check bilateral
handoffs, call linkage, evidence coverage, and exact settlement.

\noindent\textbf{Measurement protocol.}
Within each round, all schemes share the inputs and inference
configuration and run in seeded random order. We exclude warm-ups
and report per-round results, means, and sample standard deviations.
Full request processing time includes evidence persistence,
local settlement, and applicable audits. The serving runs use
signed provider receipts and audit every request under
\textsc{TokenProv} and \textsc{SignedLogPlus}.
Section~\ref{sec:eval-provider} reports separate measurements
of scoped provider proofs.

\subsection{End-to-End Overhead and Scalability}
\label{sec:evaluation-performance}
\begin{figure}
	\centering
	\includegraphics[width=\linewidth]{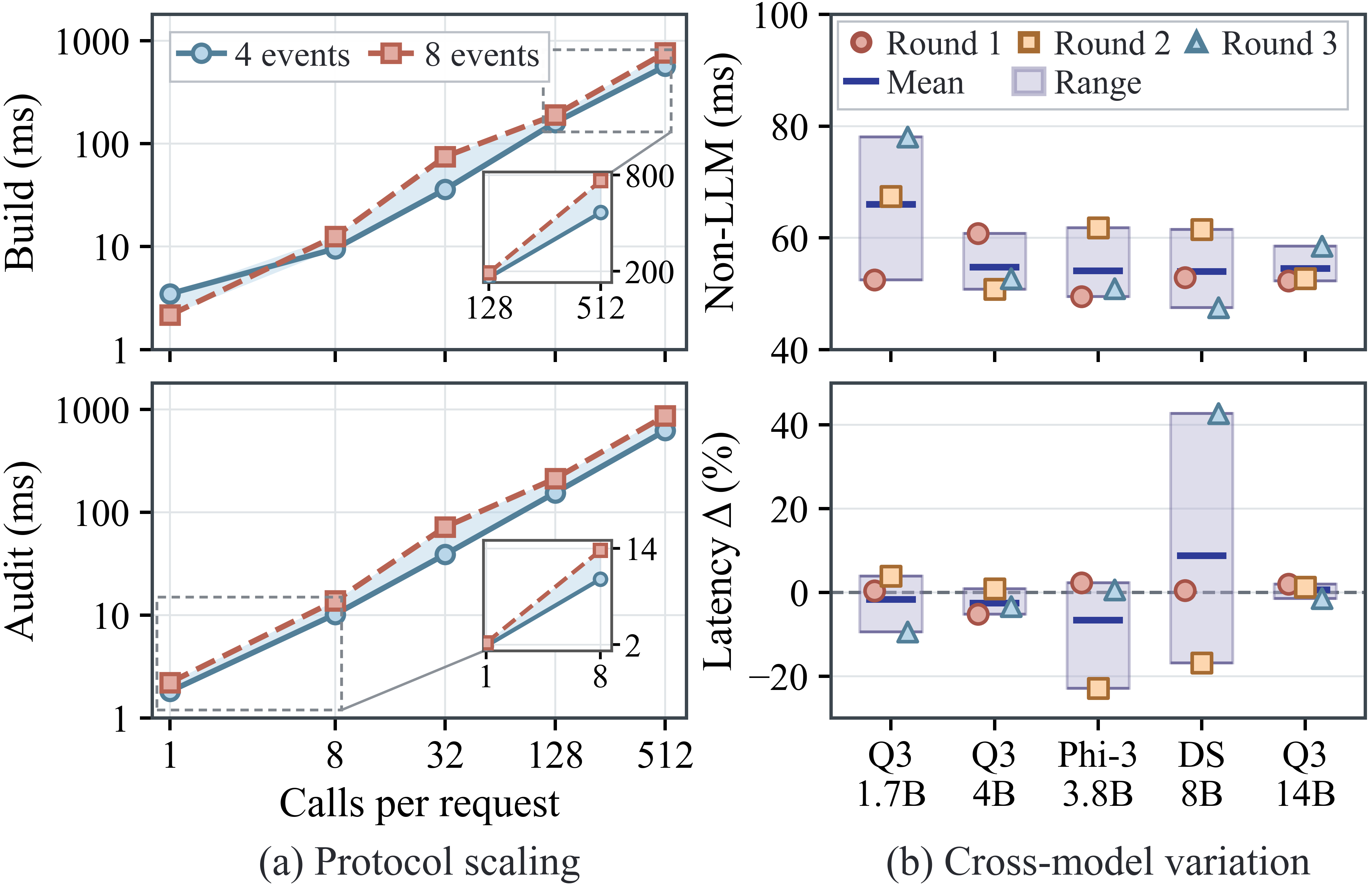}
	\caption{Protocol scalability and serving variability.
		(a) CPU construction and audit times versus calls per request.
		(b) Online non-LLM time and latency change relative to
		\textsc{Vanilla} across models.}
	\label{fig:workflow-scalability}
\end{figure}

\noindent\textbf{Cross-model serving performance.}
Across five models, \textsc{TokenProv} achieves 94.4--107.4\%
of \textsc{Vanilla}'s mean throughput over the full request
pipeline (Figure~\ref{fig:e2e-latency}). On Qwen3-1.7B, mean
request latency is 18.78\,s with \textsc{TokenProv} and 19.19\,s
with \textsc{Vanilla}. Latency varies across rounds on the shared
GPUs. For DeepSeek-R1-Distill-Llama-8B, the within-round latency
change relative to \textsc{Vanilla} averages $+8.78\%$ and reaches
$+42.71\%$ in the third round
(Figure~\ref{fig:workflow-scalability}(b)).

\begin{figure}[t]
	\centering
	\includegraphics[width=\linewidth]{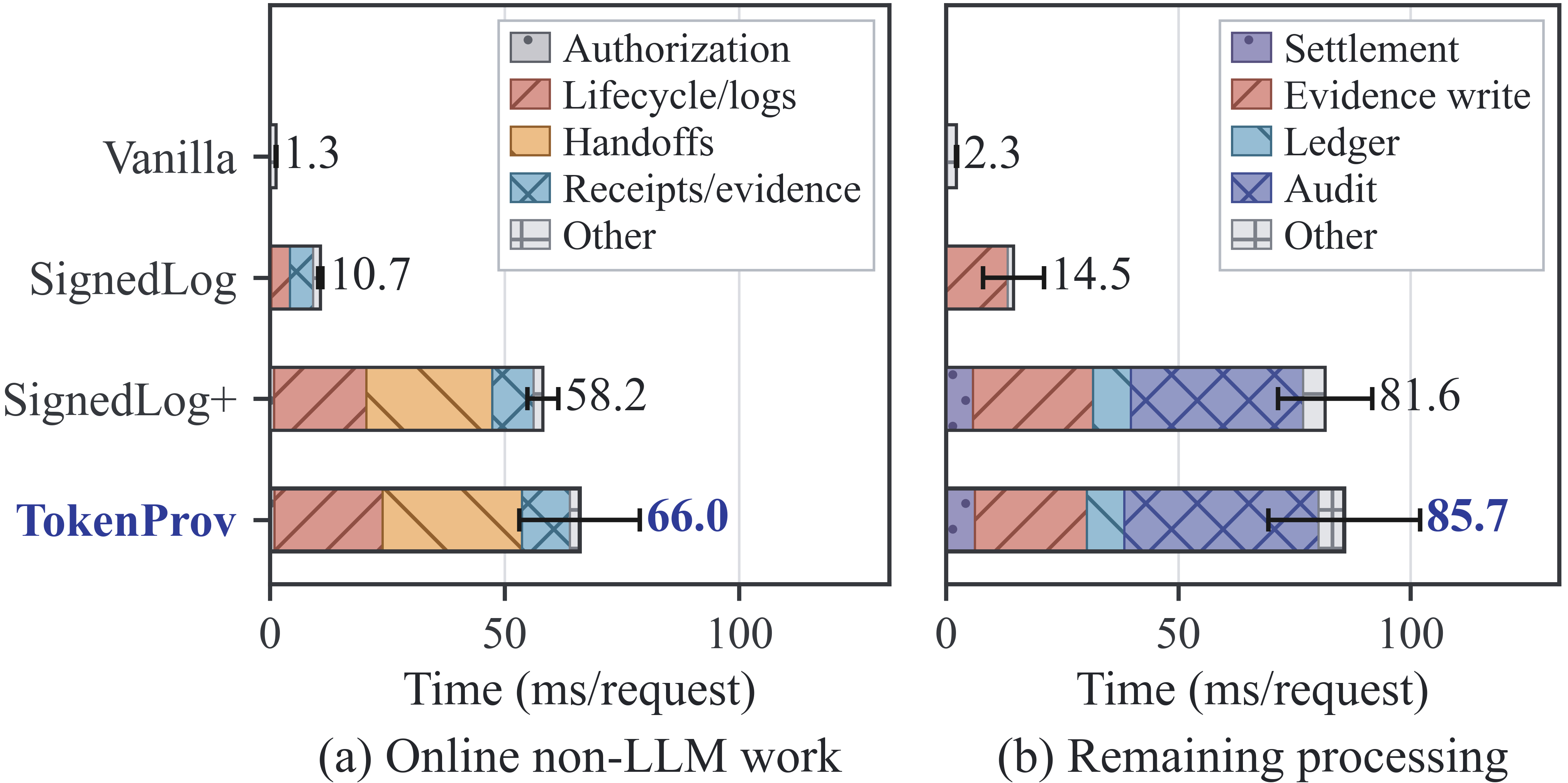}
	\caption{Non-LLM processing time on Qwen3-1.7B.
		(a) Online work. (b) Remaining request processing.}
	\label{fig:latency-breakdown}
\end{figure}

\noindent\textbf{Non-LLM processing costs.}
Across the five models, \textsc{TokenProv}'s mean online non-LLM
time ranges from 54 to 66\,ms/request
(Figure~\ref{fig:workflow-scalability}(b)). On Qwen3-1.7B,
lifecycle recording and bilateral handoffs take 23.00 and
29.70\,ms/request, respectively, together accounting for 79.9\%
of online non-LLM time (Figure~\ref{fig:latency-breakdown}).
Protocol audit and evidence persistence account for most of the
85.70\,ms/request after the online stage. Total non-LLM processing
takes 151.68\,ms/request, about 12\,ms/request more than
\textsc{SignedLogPlus}.

\noindent\textbf{Scaling with call history size.}
In the CPU microbenchmark, quadrupling the call count from
128 to 512 increases in-memory construction and audit times by
factors of 3.5--4.1 with four or eight lifecycle events per call
(Figure~\ref{fig:workflow-scalability}(a)). At 512 calls and
4,096 lifecycle events, construction and audit take 764.82
and 864.69\,ms/request, respectively.
Table~\ref{tab:b3-protocol-scaling} reports all configurations.

\subsection{Provider Assurance Overhead}
\label{sec:eval-provider}
A provider can use TEE-based attestation to avoid per-call proving
when the verifier accepts the platform's attestation and the
protected program covers the required computation. Proof-based
assurance supports deployments without a verifier-approved TEE.
We benchmark scoped Groth16 proofs by replaying retained witnesses
from 128 Qwen3-1.7B requests (1,024 calls). The circuit checks a
Poseidon accounting commitment, integer token charges, and scoped
request-hash equality. The host validates receipt-bound witnesses
and checks shared-field linkage. Appendix~\ref{app:c-assurance}
details the benchmarks.

\begin{figure}
	\centering
	\includegraphics[width=\columnwidth]{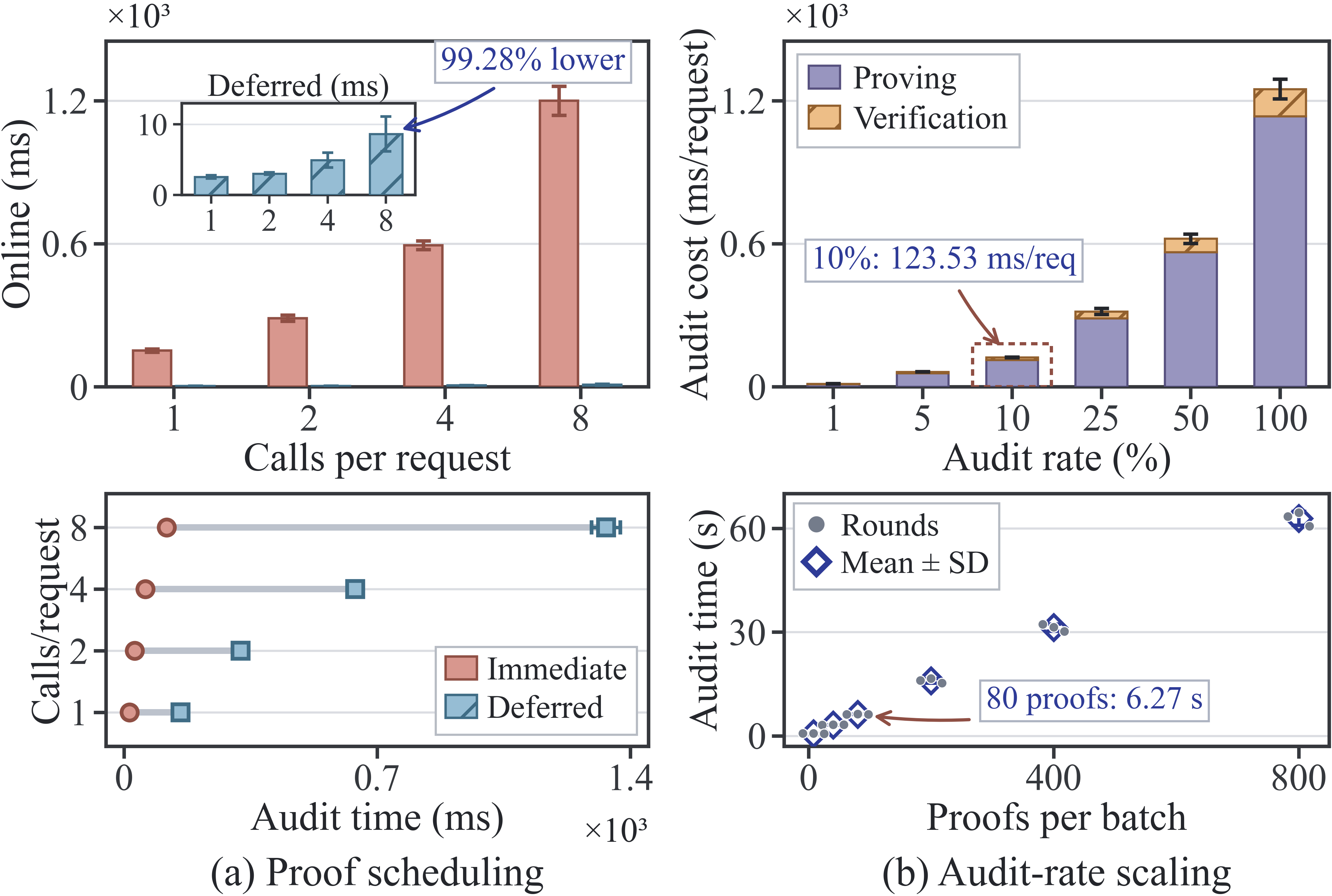}
	\caption{Proof scheduling and audit-rate scaling.
		(a) Foreground and audit times under immediate and
		deferred proving. (b) Amortized proving and verification cost
		by audit rate and elapsed audit time by proof count.}
	\label{fig:c-deferred-audit}
\end{figure}

\noindent\textbf{Deferred proving.}
With persistent workers, witness calculation and proving take
about ten times as long as verification
(Table~\ref{tab:c-assurance-components}). Both schedules persist
witnesses in the foreground and verify proofs during audit.
For an eight-call request, moving proof generation to audit cuts
foreground protocol time from 1.20\,s to 8.65\,ms, a mean paired
reduction of 99.28\% (Figure~\ref{fig:c-deferred-audit}(a)).
Total replay time is 1.32\,s with immediate proving and 1.34\,s
with deferral. Amortized proving and verification time grows
roughly in proportion to the audit rate
(Figure~\ref{fig:c-deferred-audit}(b)). At 10\%, the 80 proofs
cost 123.53\,ms/request averaged over the 100-request batch,
which two CPU workers audit in 6.27\,s.

\begin{table}
	\centering
	\caption{Assurance component costs.}
	\label{tab:c-assurance-components}
	\footnotesize
	\renewcommand{\arraystretch}{1.08}
	\begin{tabular*}{\columnwidth}{
			@{\hspace{3pt}}l@{\extracolsep{\fill}}cc@{\hspace{3pt}}
		}
		\toprule
		\textbf{Component} & \textbf{Unit} & \textbf{Mean $\pm$ SD} \\
		\midrule
		\rowcolor{gray!12}
		\multicolumn{3}{l}{\textbf{Scoped Groth16 (persistent CPU workers)}} \\
		Witness calculation + proving & ms/call & $148.45 \pm 7.64$ \\
		Verification                  & ms/call & $15.10 \pm 1.55$ \\
		JSON proof size               & B/call  & $722.98 \pm 0.47$ \\
		\midrule
		\rowcolor{gray!12}
		\multicolumn{3}{l}{\textbf{Receipt signing (laptop, ECDSA-P256)}} \\
		Software key              & ms/receipt & $0.534 \pm 0.028$ \\
		TEE-isolated key          & ms/receipt & $0.763 \pm 0.085$ \\
		Paired increment          & ms/receipt & $0.229 \pm 0.108$ \\
		\bottomrule
	\end{tabular*}
	\par\smallskip
	\parbox{\columnwidth}{\scriptsize
		Mean $\pm$ sample SD
		across three proof rounds or three paired signing sessions.}
\end{table}

\noindent\textbf{TEE-based assurance.}
We compare isolated-key and software-key receipt signing using
the same ECDSA-P256 API. In paired tests, key isolation adds
0.229\,ms per receipt on average
(Table~\ref{tab:c-assurance-components}). Receipt hashing and
accounting run on the host.

\subsection{Evidence and Settlement Overhead}
\label{sec:eval-evidence}
\begin{figure}
	\centering
	\includegraphics[width=\columnwidth]{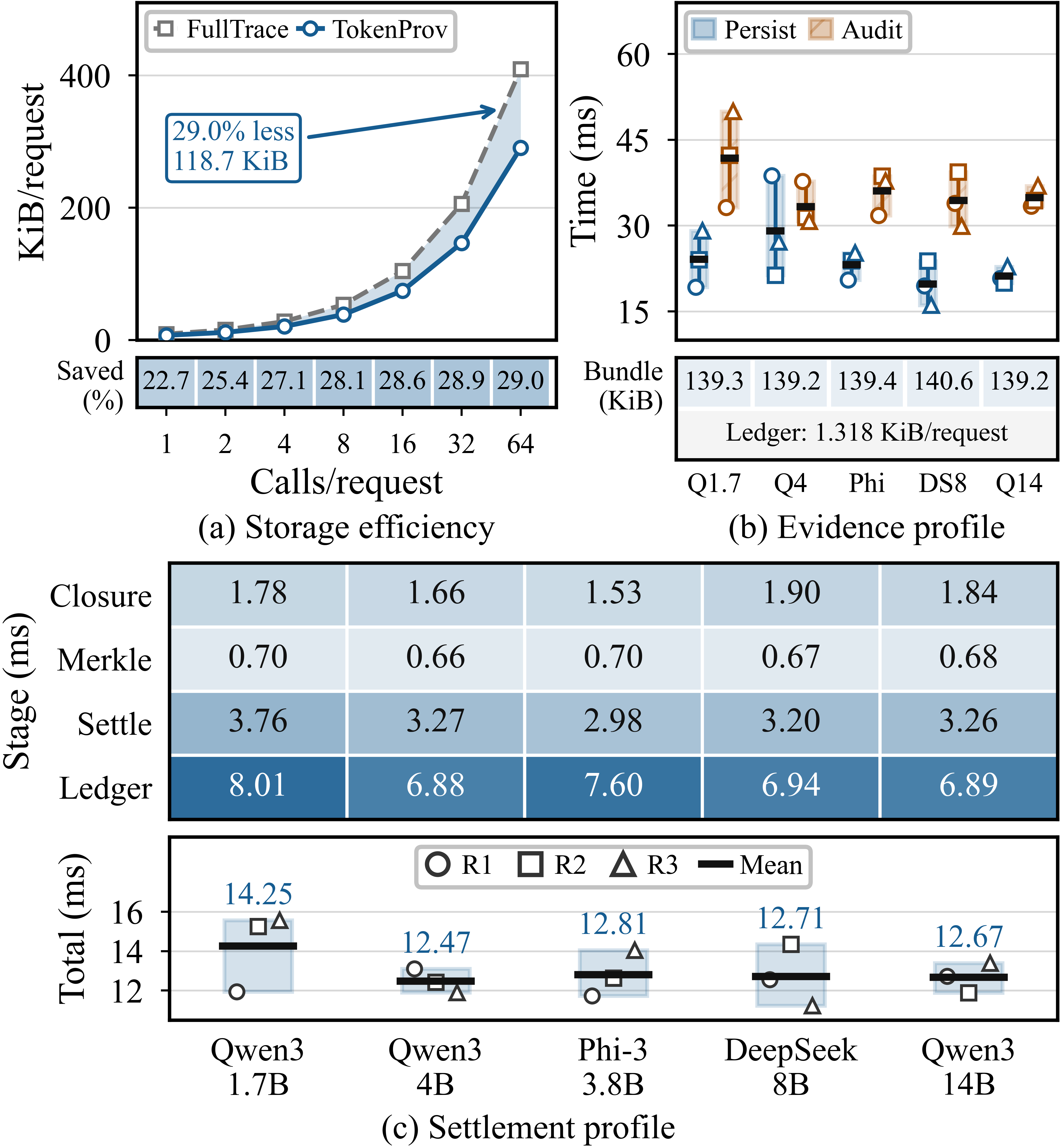}
	\caption{Evidence retention and local settlement.
		(a) Storage by call count compared with \textsc{FullTrace}.
		(b) Evidence size, persistence time, and audit time across models.
		(c) Local settlement time by stage and in total.}
	\label{fig:evidence-settlement}
\end{figure}

\noindent\textbf{Evidence retention and audit.}
In a storage microbenchmark with a common serialization format,
\textsc{TokenProv} uses 22.7--29.0\% less storage than
\textsc{FullTrace} over 1--64 calls per request.
\textsc{FullTrace} also retains available raw messages and
responses. At 64 calls, the footprint falls from 409.13 to
290.46\,KiB/request (Figure~\ref{fig:evidence-settlement}(a)).
Across the five serving models, mean evidence bundles occupy
139.16--140.59\,KiB/request, and local ledger records add
1.318\,KiB/request. Persistence takes 19.78--29.06\,ms/request,
and in-memory audit takes 33.30--41.80\,ms/request
(Figure~\ref{fig:evidence-settlement}(b)). An offline verifier
also reloads and validates the persisted bundles.

\noindent\textbf{Local settlement.}
Local settlement takes 12.47--14.25\,ms/request across models
(Figure~\ref{fig:evidence-settlement}(c)). Appending three signed
records in one local batch accounts for over half of this time.
Request closure, Merkle root construction, and settlement record
creation take 5.21--6.24\,ms/request in total, including
0.66--0.70\,ms/request for the Merkle root.
Appendix~\ref{app:evidence-settlement} reports the detailed
measurements.

\subsection{Security Effectiveness and Accountability}
\label{sec:eval-security}
\noindent\textbf{Adaptive attacks with valid signatures.}
We test aggregate-bill manipulation, duplicate or missing receipts,
omission of observable acceptances, cross-call evidence splicing,
and unauthorized deployment substitution. The adversary can re-sign
records for $H$ and $B$, and for $P$ in the splicing and substitution
tests. We fix the user authorization and independently retained
handoffs, check each attack query's signatures, and confirm its
violation against an independent reference. Each episode uses
first-failure feedback and ends at the first accepted violation
or after 16 audited queries. \textsc{SignedLog} accepts the first
query in all 100 episodes. \textsc{SignedLogPlus} and
\textsc{TokenProv} each reject all 1,600 attack queries. All three
schemes accept their 100 benign controls
(Table~\ref{tab:e-security-accountability}(a)).

\noindent\textbf{Cross-request evidence binding.}
A separate microbenchmark uses genuine call, receipt, and assurance
records across requests with the same call label. Source-context
checks pass in all 1,500 trials. \textsc{TokenProv} rejects every
transplant at the request-binding check in
$\mathcal{R}_{\mathsf{prov}}$, including pairs with the same model
and deployment (Table~\ref{tab:e-security-accountability}(b)).

\noindent\textbf{First-failure localization.}
The first failed relation matches the injected stage in all
2,500 trials. Responsibility intervals span zero to two role
boundaries, from $[H,H]$ in the lifecycle fixture to $[H,P]$
in the handoff fixture. Mean audit time ranges from 0.168 to
14.350\,ms across the five stages
(Table~\ref{tab:e-security-accountability}(c)).

\noindent\textbf{Accountability under collusion.}
In the $B{+}P$ control, a retained signed $H\!\to\!B$ handoff
contradicts the submitted provider evidence. The auditor rejects
all 500 trials and reports $[B,P]$. The $H{+}B$ hidden-execution
and $H{+}B{+}P$ consistent-transcript controls each yield 500/500
acceptances with internally consistent audit evidence
(Table~\ref{tab:e-security-supplement}(b)).

\begin{table}
	\centering
	\caption{Transcript audit outcomes and fault localization.}
	\label{tab:e-security-accountability}
	\footnotesize
	\setlength{\tabcolsep}{3pt}
	\renewcommand{\arraystretch}{1.08}
	
	\begin{tabularx}{\columnwidth}{
			@{}
			>{\raggedright\arraybackslash}p{0.23\columnwidth}
			>{\centering\arraybackslash}p{0.27\columnwidth}
			>{\centering\arraybackslash}p{0.20\columnwidth}
			>{\centering\arraybackslash}X
			@{}
		}
		\toprule
		\rowcolor{black!7}[0pt][0pt]
		\multicolumn{4}{@{}l@{}}{
			\hspace{.5em}\textbf{(a) Adaptive attacks with valid signatures}
		} \\
		{\scriptsize\textbf{Scheme}}
		& {\scriptsize\textbf{Successful episodes}}
		& {\scriptsize\textbf{Attack queries}}
		& {\scriptsize\textbf{Benign rejections}} \\
		\midrule
		\textsc{SignedLog}     & 100/100        & 100   & 0/100 \\
		\textsc{SignedLogPlus} & \textbf{0/100} & 1,600 & 0/100 \\
		\textsc{TokenProv}     & \textbf{0/100} & 1,600 & 0/100 \\
		\bottomrule
	\end{tabularx}
	
	\par\medskip
	\begin{tabularx}{\columnwidth}{
			@{}
			>{\raggedright\arraybackslash}p{0.52\columnwidth}
			>{\centering\arraybackslash}p{0.27\columnwidth}
			>{\centering\arraybackslash}X
			@{}
		}
		\toprule
		\rowcolor{black!7}[0pt][0pt]
		\multicolumn{3}{@{}l@{}}{
			\hspace{.5em}\textbf{(b) Cross-request splicing}
			\enspace (accepted/tested)
		} \\
		{\scriptsize\textbf{Source--target pair}}
		& {\scriptsize\textbf{Source-local checks}}
		& {\scriptsize\textbf{\textsc{TokenProv}}} \\
		\midrule
		Same model and deployment
		& 500/500 & \textbf{0/500} \\
		Same model, different deployment
		& 500/500 & \textbf{0/500} \\
		Different model and deployment
		& 500/500 & \textbf{0/500} \\
		\bottomrule
	\end{tabularx}
	
	\par\medskip
	\setlength{\tabcolsep}{3pt}
	\begin{tabular*}{\columnwidth}{@{\extracolsep{\fill}}lcccr@{}}
		\toprule
		\rowcolor{black!7}[0pt][0pt]
		\multicolumn{5}{@{}l@{}}{
			\hspace{.5em}\textbf{(c) First-failure localization}
		} \\
		
		\multirow[c]{2}{*}{\textbf{Injected stage}}
		& \textbf{First failed}
		& \textbf{Stage}
		& \textbf{Interval}
		& \multicolumn{1}{c}{\textbf{Mean audit}} \\
		
		& \textbf{relation}
		& \textbf{matches}
		& (hops)
		& \multicolumn{1}{c}{(ms)} \\
		
		\midrule
		Authorization & $\mathcal{R}_{\mathsf{auth}}$
		& 500/500 & $[U,H]$ (1) & 0.168 \\
		Lifecycle & $\mathcal{R}_{\mathsf{life}}$
		& 500/500 & $[H,H]$ (0) & 3.476 \\
		Handoff & $\mathcal{R}_{\mathsf{hand}}$
		& 500/500 & $[H,P]$ (2) & 3.430 \\
		Provider & $\mathcal{R}_{\mathsf{prov}}$
		& 500/500 & $[B,P]$ (1) & 9.165 \\
		Settlement & $\mathcal{R}_{\mathsf{set}}$
		& 500/500 & $[H,B]$ (1) & 14.350 \\
		\bottomrule
	\end{tabular*}

	\par\smallskip
	\parbox{\columnwidth}{\scriptsize
		Panels (b) and (c) report earlier microbenchmarks with five
		fixtures per configuration and 100 repetitions per fixture. Interval widths count hops along
		$U\!\rightarrow\!H\!\rightarrow\!B\!\rightarrow\!P$.}
\end{table}

\subsection{Ablation Study}
\label{sec:eval-ablation}
\noindent\textbf{First-failure rejection rules.}
Each control changes rejection to acceptance if the first failed
relation matches the selected rule. For each of the four cases in
Table~\ref{tab:security-ablation}, the unmodified auditor rejects
all 500 evaluations, and the matched override accepts all 500.

\begin{table}[!t]
	\centering
	\caption{Audit acceptance under rule overrides.}
	\label{tab:security-ablation}
	\footnotesize
	\setlength{\tabcolsep}{3pt}
	\renewcommand{\arraystretch}{1.10}
	\begin{tabularx}{\columnwidth}{
			@{\hspace{4pt}}
			>{\centering\arraybackslash}p{0.18\columnwidth}
			>{\raggedright\arraybackslash}X
			cc
			@{\hspace{4pt}}
		}
		\toprule
		\textbf{Target rule} & \textbf{Injected violation}
		& \textbf{Unmodified} & \textbf{Override} \\
		\midrule
		$\mathcal{R}_{\mathsf{life}}$ & Unanchored call insertion
		& \textbf{0/500} & 500/500 \\
		$\mathcal{R}_{\mathsf{hand}}$ & $H\!\to\!B$ handoff tampering
		& \textbf{0/500} & 500/500 \\
		$\mathcal{R}_{\mathsf{prov}}$ & Cross-call evidence splicing
		& \textbf{0/500} & 500/500 \\
		$\mathcal{R}_{\mathsf{set}}$ & Aggregate-bill manipulation
		& \textbf{0/500} & 500/500 \\
		\bottomrule
	\end{tabularx}
	\par\smallskip
	{\scriptsize\raggedright\noindent
		Five fixtures per condition, with 100 evaluations per fixture.\par}
\end{table}

\begin{figure}[!t]
	\centering
	\includegraphics[width=0.9\columnwidth]{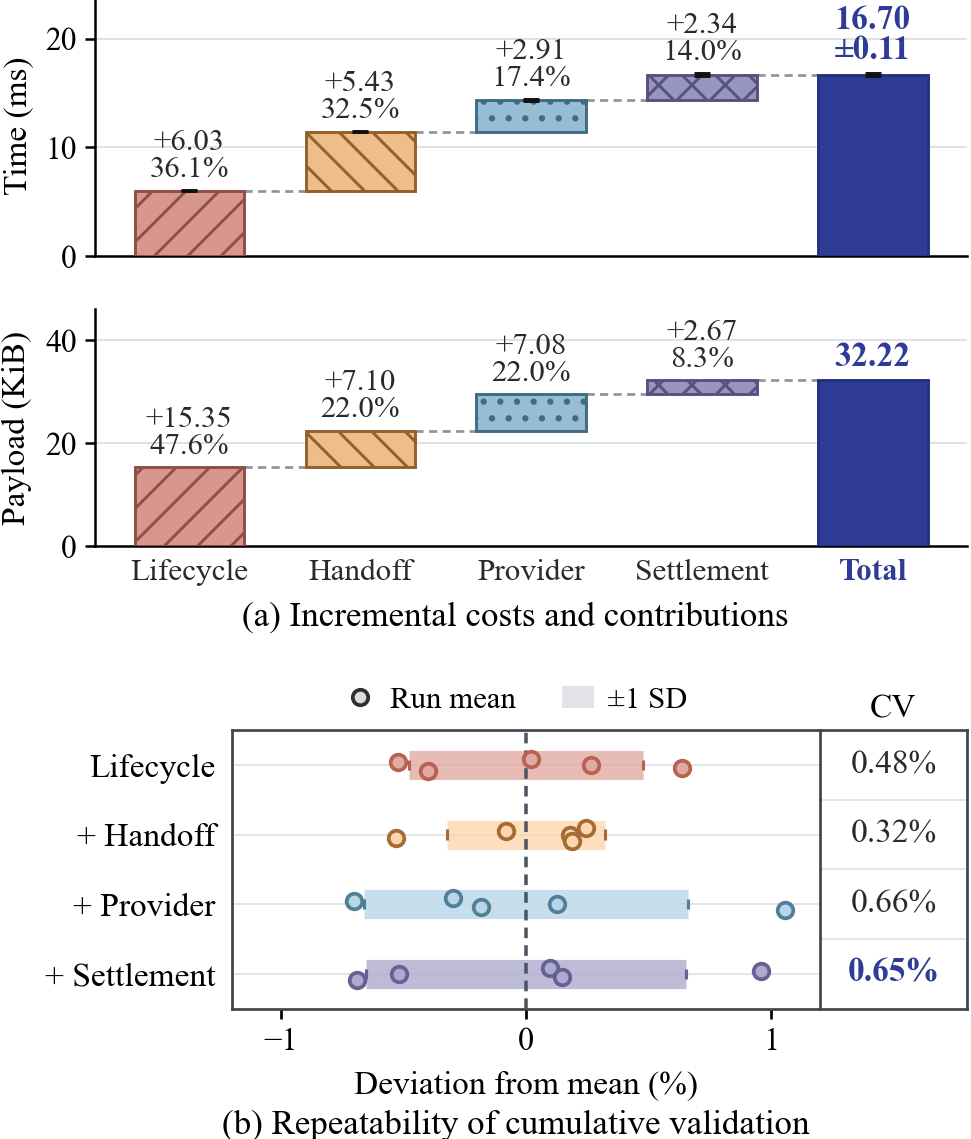}
	\caption{Cumulative validation costs and repeatability.
		(a) Increments in wall time and selected payload size.
		(b) Run-to-run variation in cumulative wall time.}
	\label{fig:ablation-cost}
\end{figure}

\noindent\textbf{Incremental validation costs.}
On prebuilt fixtures with eight calls and 32 lifecycle events,
we cumulatively enable lifecycle, bilateral handoff, provider,
and settlement checks. Validation and payload-field selection
take $16.70\pm0.11$\,ms per fixture after all four stages are
enabled (mean $\pm$ sample SD across five runs). The selected
fields encode to 32.22\,KiB. In this activation order, lifecycle
and handoff checks account for 68.6\% of the measured time and
69.7\% of the selected payload (Figure~\ref{fig:ablation-cost}(a)).
The coefficient of variation across runs ranges from 0.32\% to
0.66\% over the four configurations with checks enabled
(Figure~\ref{fig:ablation-cost}(b)).
Appendix~\ref{app:f-ablation} details the controls, timing scope,
and run-level results.

\section{Related Work}
\label{sec:related-work}
\subsection{Agent Provenance and Accountability}

Security evaluations of LLM agents characterize failures arising
from tool use and untrusted inputs
~\cite{ruan2024toolemu,debenedetti2024agentdojo}.
PROV-AGENT extends W3C PROV to capture agent interactions within
workflow provenance~\cite{souza2025provagent}. Work on authenticated
delegation scopes the authority granted to agents
~\cite{south2025delegation}, and Authenticated Workflows uses
cryptographic attestations to establish workflow dependencies
under context-sensitive policies~\cite{rajagopalan2026authenticated}.

PeerReview checks distributed execution against a reference
protocol, and CATS provides accountability for network storage
~\cite{haeberlen2007peerreview}.
In-toto attests software supply-chain steps, LPM captures
kernel-level provenance, and ALASTOR reconstructs serverless
intrusions~\cite{torresarias2019intoto,bates2015lpm,datta2022alastor}.
\textsc{TokenProv} uses authenticated call histories and bilateral
handoffs to carry signed user authorization through runtime call
creation to request-level settlement.

\subsection{Verifiable and Attested Inference}
Verifiable-computation systems support outsourced ML,
including efficient proofs for ML inference operators
~\cite{chen2024zkml,hao2024zkmlnonlinear} and specialized proofs
for LLM inference~\cite{sun2024zkllm,qu2025zkgpt}.
VeriML uses verification of sampled training iterations to
support resource accounting and fair payment
~\cite{zhao2021veriml}.

Slalom combines protected execution with integrity checks for
outsourced accelerator operations, Graviton provides trusted GPU
execution, and SAGE supports software-based GPU attestation
~\cite{tramer2019slalom,volos2018graviton,ivanov2023sage}.
State-continuity mechanisms address rollback in protected
execution~\cite{parno2011memoir,niu2022narrator}.
AFTUNE retrospectively audits proprietary fine-tuning and
inference by checking sampled execution blocks inside a TEE
~\cite{jin2026aftune}. \textsc{TokenProv} defines a provider
relation realizable with proofs or attested state transitions.
The relation binds the accepted deployment, cache-dependent
execution, metered usage, and charge to the same invocation.

\subsection{Auditing LLM Services and Token Usage}

Audits of opaque LLM services examine model substitution
~\cite{cai2025modelaudit}, cross-user prompt caching
~\cite{gu2025promptcache}, and hidden reasoning-token usage
~\cite{sun2025coin}. Studies of pay-per-token
pricing examine incentives and methods for overcharging
~\cite{artola2025overcharging,hoque2026tokeninflation}.
Sequential tests detect usage misreporting across queries
~\cite{velasco2026auditing}, and a complementary position paper
proposes a layered architecture for execution verification,
secure logging, and user-facing audits~\cite{sun2025invisible}.
IMMACULATE uses verifiable computation to audit inference
and token accounting on sampled requests~\cite{guo2026immaculate}.

At the gateway layer, GateScope measures behavioral and billing
discrepancies~\cite{lin2026gateways}. The Attested Gateway Runtime
(AGR) signs evidence binding routing decisions, upstream
endpoints, and response streams~\cite{wang2026gateway}.
AEX attests request-to-output lineage across API hops, including
intermediary transformations~\cite{guan2026aex}. InterSAGE
proposes signed usage records that attribute token consumption
to agent identities and delegation chains~\cite{zou2026intersage}.

\textsc{TokenProv} reconciles each audited request's bill with
its authenticated call history. The audit matches observable
provider acceptances to settled receipts one-to-one and
recomputes the bill using execution-bound usage and the
applicable pricing rules.
		
\section{Conclusion}
\label{sec:conclusion}

\textsc{TokenProv} makes request-level LLM bills auditable against
signed user authorization and authenticated call evidence. The audit
links each settled token charge to its execution and matches
observable provider acceptances to receipts one-to-one. We prove
audit soundness relative to authenticated evidence.
Across five models, serving with signed provider receipts takes
54--66\,ms of online non-LLM processing and 12.47--14.25\,ms of
local settlement per request. In an eight-call replay, moving
scoped proof generation to the audit phase reduces foreground
protocol time from 1.20\,s to 8.65\,ms. The auditor rejects all
1,600 tested adaptive attack queries with valid signatures.

\section*{Acknowledgment}
This work was supported in part by the Beijing Natural Science
Foundation under Grants L2601001 and QY26354, the Science and
Technology Projects of Xizang Autonomous Region under Grant
XZ202401ZY0027, the Beijing Advanced Innovation Center for Future
Blockchain and Privacy Computing, and the Beijing Key Laboratory
of Multimodal Data Intelligent Perception and Governance.
This work is partially supported by Ripple under University Blockchain Research Initiative (UBRI)\cite{fund}.

\bibliographystyle{IEEEtran}
\bibliography{sample-base}
		
\appendices

\section{Notation Summary}
\label{app:notation-summary}

Figure~\ref{fig:entity-notation} summarizes the notation used in
the service model, protocol, and audit.

\begin{figure*}[t]
	\centering
	\footnotesize
	\setlength{\fboxrule}{0.35pt}
	\setlength{\fboxsep}{0pt}
	\renewcommand{\arraystretch}{1.08}
	
	\newsavebox{\tpNotationUser}
	\newsavebox{\tpNotationHarness}
	\newsavebox{\tpNotationBroker}
	\newsavebox{\tpNotationProvider}
	\newsavebox{\tpNotationOther}
	\newdimen\tpNotationHeight
	
	\newcommand{\tpNotationPrepare}[4]{%
		\sbox{#1}{%
			\begin{minipage}[t]{\dimexpr#2-2\fboxrule\relax}
				\setlength{\fboxsep}{4pt}%
				\noindent\colorbox{gray!10}{%
					\parbox{\dimexpr\linewidth-2\fboxsep\relax}{%
						\strut\textbf{#3}}}%
				\par\vspace{3pt}%
				\noindent\hspace*{\fboxsep}%
				\begin{tabularx}{\dimexpr\linewidth-2\fboxsep\relax}
					{@{}l@{\hspace{.7em}}>{\raggedright\arraybackslash}X@{}}
					#4
				\end{tabularx}%
				\par\vspace{3pt}%
		\end{minipage}}%
	}
	\newcommand{\tpNotationMeasure}[1]{%
		\ifdim\dimexpr\ht#1+\dp#1\relax>\tpNotationHeight
		\tpNotationHeight=\dimexpr\ht#1+\dp#1\relax
		\fi
	}
	\newcommand{\tpNotationDraw}[1]{%
		\fcolorbox{black!38}{white}{%
			\parbox[t][\tpNotationHeight][t]{\wd#1}{%
				\noindent\usebox{#1}}}%
	}
	\newcommand{\tpNotationTopWidth}{\dimexpr(\linewidth-10pt)/3\relax}
	\newcommand{\tpNotationBottomWidth}{\dimexpr(\linewidth-5pt)/2\relax}
	
	\tpNotationPrepare{\tpNotationUser}{\tpNotationTopWidth}{User $U$}{%
		$r$ & Request ID \\
		$C_r^{\mathsf{in}}$ & Private-input commitment \\
		$\Psi_r$ & Execution and billing policy \\
		$\mathsf{Auth}_r,h_r^{\mathsf{auth}}$ & Signed authorization, digest \\
		$\mathsf{ans}_r^U,B_r^U$ & Retained response and charge \\
	}
	
	\tpNotationPrepare{\tpNotationHarness}{\tpNotationTopWidth}{Harness $H$}{%
		$\mathsf{cid}_i$ & Call-attempt ID \\
		$\mathcal{Q}_r$ & Generated attempt IDs \\
		$\iota_i,\vartheta_i$ & Call intent and constraints \\
		$\mathcal{G}_r^{\mathsf{life}}$ & Authenticated lifecycle DAG \\
		$\mathsf{ans}_r,C_r^{\mathsf{ans}}$ & Final response, commitment \\
		$\mathsf{Close}_r$ & Signed closure record \\
	}
	
	\tpNotationPrepare{\tpNotationBroker}{\tpNotationTopWidth}{Broker $B$}{%
		$\rho_i,\theta_i$ & Transform rule and serving configuration \\
		$q_i$ & Provider-facing call \\
		$\mathcal{F}_r$ & Auxiliary fee multiset \\
		$B_r$ & Broker-reported request charge \\
		$\mathsf{Settle}_r$ & Settlement signed by $B$ and $H$ \\
	}
	
	\tpNotationPrepare{\tpNotationProvider}{\tpNotationBottomWidth}{Provider $P$}{%
		$\mathsf{Reg}_j,\Gamma_j$ & Signed registration and public descriptor \\
		$\mathcal{Q}_r^P$ & Provider-accepted call IDs \\
		$\boldsymbol{\tau}_i^{\mathsf{in}},\boldsymbol{\tau}_i^{\mathsf{out}}$
		& Logical input and returned token sequences \\
		$\kappa_i^-,\kappa_i^+,A_i^\kappa$
		& Logical KV-cache states and pre-state evidence \\
		$\phi_i,\mathbf{u}_i,b_i$ & Termination code, metered usage, and call charge \\
		$C_i^{\mathsf{res}}$ & Result commitment \\
		$W_i^{\mathsf{call}},C_i^W$ & Call witness and commitment \\
		$\mathsf{Rcpt}_i,\mathsf{nf}_i$ & Signed receipt and call nullifier \\
		$\alpha_i^{\mathsf{assr}},\eta_i^{\mathsf{assr}}$
		& Assurance anchor and artifact \\
		$\mathcal{R}_{\mathsf{P}}^i$ & Call-level assurance relation \\
	}
	
	\tpNotationPrepare{\tpNotationOther}{\tpNotationBottomWidth}{Other notation}{%
		$L,V$ & Append-only ledger and off-chain evidence store \\
		$\mathsf{Hnd}_i^{X\rightarrow Y}$ & Bilateral handoff from $X$ to $Y$ \\
		$\mathcal{Q}_r^{\mathsf{obs}},\mathcal{Q}_r^{P,\mathsf{obs}}$
		& Observable attempt and provider-accepted call IDs \\
		$\mathcal{Q}_r^{\mathsf{bill}}$ & Billable accepted-call IDs \\
		$\mathcal{S}_r^{\mathsf{rcpt}}$ & Settled receipt multiset \\
		$B_r^\star$ & Recomputed request charge \\
		$\mathcal{T}_r,\mathcal{W}_r$ & Audit transcript and private evidence \\
		$\mathcal{R}_{\mathsf{TP}},\delta_r$ & Audit validity relation and decision bit \\
		$\mathcal{C}_r,\mathsf{Resp}_r$
		& Candidate fault positions and responsibility interval \\
		$\mathsf{Leak}_r$ & Permitted audit leakage \\
	}
	
	\tpNotationHeight=0pt
	\tpNotationMeasure{\tpNotationUser}
	\tpNotationMeasure{\tpNotationHarness}
	\tpNotationMeasure{\tpNotationBroker}
	\noindent
	\tpNotationDraw{\tpNotationUser}\hfill
	\tpNotationDraw{\tpNotationHarness}\hfill
	\tpNotationDraw{\tpNotationBroker}\par\vspace{5pt}
	
	\tpNotationHeight=0pt
	\tpNotationMeasure{\tpNotationProvider}
	\tpNotationMeasure{\tpNotationOther}
	\noindent
	\tpNotationDraw{\tpNotationProvider}\hfill
	\tpNotationDraw{\tpNotationOther}
	
	\caption{Principal notation in \textsc{TokenProv}.}
	\label{fig:entity-notation}
\end{figure*}

\section{Security Proofs}
\label{app:security-proofs}

The integrity claims are relative to $\mathcal{T}_r$ under the
assumptions of Section~\ref{sec:threat-model}.

\subsection{Authenticated Structural Integrity}
\label{app:structural-integrity}

\begin{proof}[Proof of Lemma~\ref{lem:structural-integrity}]
	Fix an accepting audit. The verified $\mathsf{Auth}_r$
	fixes $(r,h_r^{\mathsf{auth}})$ by
	\eqref{eq:authorization-digest}.
	$\mathcal{R}_{\mathsf{life}}=1$ gives an acyclic lifecycle
	rooted in that authorization, with one $H$-signed spawn and,
	by \eqref{eq:unique-terminal-event}, one terminal event per
	observable call. $\mathcal{R}_{\mathsf{hand}}=1$ verifies the
	bilateral signatures, boundary openings, predecessor
	references, and acceptance-context uniqueness
	under \eqref{eq:bilateral-handoff}.
	
	An accepted signature on a new message under an honest key
	violates unforgeability. Distinct canonical records under
	one digest give a hash collision, and distinct values opened
	from one commitment violate binding.
	
	Finalized $\mathsf{Settle}_r$ binds
	$\mathsf{root}_r^{\mathsf{evid}}$ directly and
	$\mathsf{root}_r^{\mathsf{life}}$ through
	$\mathsf{Hash}(\mathsf{Close}_r)$.
	Because the Merkle construction authenticates length and leaf
	positions, distinct canonical sequences opening either root
	yield a hash collision.
\end{proof}

\subsection{Provider-Assurance Soundness}
\label{app:proof-provider-soundness}

\begin{proof}[Proof of Lemma~\ref{lem:provider-soundness}]
	Fix an adaptively designated
	$\mathsf{cid}_i\in\mathcal{Q}_r^{P,\mathsf{obs}}$.
	The verifier derives $\Psi_r$ from $\mathsf{Auth}_r$ and
	reconstructs the call context from the authenticated
	registration, handoffs, receipt, and cache predecessor.
	The authorized profile fixes the mode and verifier for
	\eqref{eq:provider-composite-relation}.
	
	\emph{Proof-based realization.}
	The verifier uses $C_i^W=\alpha_i^{\mathsf{assr}}$ and
	the reconstructed $x_i^{\mathsf{zk}}$. Knowledge soundness
	yields $(W_i^{\mathsf{call}},d_i^W)$ satisfying
	\eqref{eq:cp-relation} for $\mathcal{R}_{\mathsf{P}}^i$.
	
	\emph{Attested realization.}
	The receipt binds $\eta_i^{\mathsf{assr}}$ through
	$\alpha_i^{\mathsf{assr}}$. Quote verification and
	\eqref{eq:attested-transition-check} authenticate the
	registered program's transition on $(x_i^T,y_i^T)$ from
	an admissible predecessor state. Soundness of any required
	external proof gives a satisfying completion for those same
	boundary values. Together they satisfy
	$\mathcal{R}_{\mathsf{P}}^i$.
	
	Thus an accepting provider check implies the existence of a
	call witness except with negligible probability, giving
	\eqref{eq:provider-soundness}.
\end{proof}

\subsection{Settlement and End-to-End Soundness}
\label{app:proof-end-to-end-soundness}

\begin{proof}[Proof of Lemma~\ref{lem:settlement-exactness}]
	The finalized evidence root binds the receipt multiset
	$\mathcal{S}_r^{\mathsf{rcpt}}$ with multiplicity.
	$\mathcal{R}_{\mathsf{cover}}=1$ gives
	\[
	\forall\,c,\qquad
	N_r(c)=\mathbf{1}[c\in\mathcal{Q}_r^{P,\mathsf{obs}}],
	\]
	including nonbillable acceptances.
	$\mathcal{R}_{\mathsf{uniq}}$ enforces receipt and fee
	uniqueness, and $L$ enforces initial-settlement uniqueness.
	
	By Lemma~\ref{lem:provider-soundness}, the matched receipts
	substantiate $(\phi_i,b_i)$ except with negligible probability.
	The audit applies $\Psi_r$ to derive the billable set and
	validate the fees, then recomputes $B_r^\star$ by
	\eqref{eq:request-bill}. Acceptance requires
	$B_r^U=B_r=B_r^\star$.
\end{proof}

\begin{proof}[Proof of Theorem~\ref{thm:audit-soundness}]
	Let $\mathsf{Bad}_{\mathsf{bind}}$ denote an accepted
	forgery, an exhibited hash collision, or a commitment-binding
	violation. Let $\mathsf{Bad}_{\mathsf{CP}}$ denote joint
	extraction failure for the accepted proof-based artifacts in
	this audit, and $\mathsf{Bad}_{\mathsf{tee}}$ a failure of
	the attested-transition or external-proof guarantee for any
	accepted attested artifact.
	
	Outside these events, each observable provider acceptance
	has a call witness $w_i$ from joint extraction or from an
	authenticated transition and any required external completion.
	A disagreement on a shared commitment $C$ gives
	\[
	\mathsf{Open}(pp,C,z,d)
	=\mathsf{Open}(pp,C,z',d')=1,\qquad z\ne z',
	\]
	which violates binding. A disagreement under a shared digest
	gives a hash collision. If $A_i^\kappa$ references the
	verified post-state $\kappa_\ell^+$, the predecessor check
	enforces $\kappa_i^-=\kappa_\ell^+$ for the referenced
	cache object, deployment, and cache-rule version.
	
	Consequently,
	$
	\mathcal{W}
	=\bigl(\mathcal{W}_r,(w_i)_{\mathsf{cid}_i
		\in\mathcal{Q}_r^{P,\mathsf{obs}}}\bigr)
	$
	satisfies $\mathcal{R}_{\mathsf{prov}}$ on the same records
	for which the audit checks
	$\mathcal{R}_{\mathsf{auth}}$, $\mathcal{R}_{\mathsf{life}}$,
	$\mathcal{R}_{\mathsf{hand}}$, and $\mathcal{R}_{\mathsf{set}}$.
	Hence $\mathcal{W}$ satisfies \eqref{eq:tokenprov-validity}.
	The primitive assumptions and joint knowledge soundness bound
	the first two bad events. A polynomial union bound over
	attested checks, using Lemma~\ref{lem:provider-soundness},
	bounds the third. For $\mathsf{Fail}_r$ in
	\[
	\begin{aligned}
		\Pr[\mathsf{Fail}_r]
		&\leq \Pr[\mathsf{Bad}_{\mathsf{bind}}]
		+ \Pr[\mathsf{Bad}_{\mathsf{CP}}]
		+ \Pr[\mathsf{Bad}_{\mathsf{tee}}] \\
		&\leq \mathsf{negl}(\lambda).
	\end{aligned}
	\]
\end{proof}

\subsection{Deferred Proving and Localization}
\label{app:proof-deferred}

\begin{proof}[Proof of Proposition~\ref{prop:deferred-proof}]
	Finalization fixes $C_i^W$ and its registered parameters
	before audit selection or challenge. Joint extraction within a single audit yields a valid opening of $C_i^W$ for each accepted deferred proof for that receipt. Two distinct extracted witness values give a binding violation. 
	A PPT reduction can find any disagreeing pair among the extracted openings.
\end{proof}

\noindent\textbf{Localization coverage.}
For a reproducible failure, let $S_r$ be the positions remaining
after exclusions justified by the authenticated inputs to the
first failed check. Every position in $\mathcal{C}_r$ has an
execution consistent with the retained evidence and thus survives
exclusion. For $\mathcal{C}_r\ne\varnothing$,
\[
\mathcal{C}_r\subseteq S_r
\subseteq\{\min S_r,\ldots,\max S_r\},
\]
which gives \eqref{eq:localization-coverage}. With no justified
exclusions, $S_r=[4]$ gives $[U,P]$.

\subsection{Audit Confidentiality}
\label{app:proof-confidentiality}

\begin{proof}[Proof of Theorem~\ref{thm:audit-confidentiality}]
	From $\mathsf{Leak}_r$, the simulator restores the
	adversary's pre-audit state and random tape and replays
	the ordered observations $o_1,\ldots,o_n$ verbatim.
	Let $v_t^b$ contain the adversary's state and view prefix
	after $t$ observations in execution
	$b\in\{\mathsf{real},\mathsf{sim}\}$. With the tape fixed,
	\[
	\begin{aligned}
		v_0^{\mathsf{real}}=v_0^{\mathsf{sim}},\quad
		v_{t+1}^{b}=F_{\mathcal A}(v_t^b,o_{t+1}).
	\end{aligned}
	\]
	Induction gives identical views after $n$ observations.
	Replay takes polynomial time and establishes
	\eqref{eq:audit-confidentiality}.
\end{proof}

\section{Additional Evaluation Results}
\label{app:additional-evaluation}
\begin{figure}[!t]
	\centering
	\includegraphics[width=0.96\columnwidth]
	{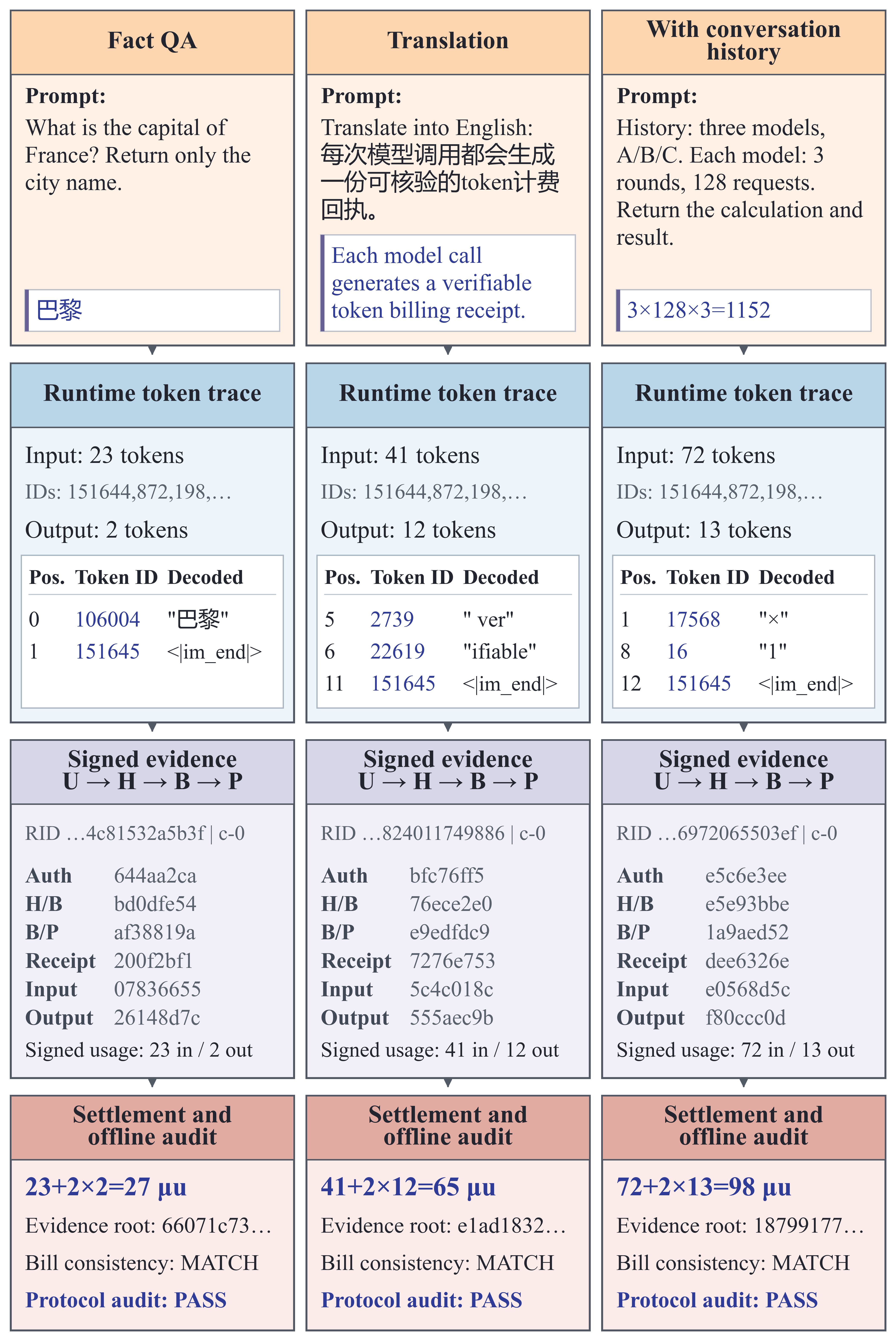}
	\caption{Token provenance for three example requests.}
	\label{fig:tokenprovenancecaseboxes}
\end{figure}

Figure~\ref{fig:tokenprovenancecaseboxes} traces three example
requests. Parentheses in the tables give sample standard deviations.

\begingroup
\definecolor{tpSDblue}{HTML}{2D5873}
\newcommand{\tpstat}[2]{%
	\mbox{#1\kern.12em\raisebox{-.24ex}{%
			\textcolor{tpSDblue}{\fontsize{6.2}{7}\selectfont (#2)}}}%
}
\setlength{\tabcolsep}{2pt}
\renewcommand{\arraystretch}{0.98}

\newenvironment{tpapptable}[1]{%
	\par\smallskip\noindent
	\begin{minipage}{\columnwidth}
		\setlength{\abovecaptionskip}{3pt}%
		\centering\captionof{table}{#1}\footnotesize
	}{\end{minipage}\par\smallskip}

\subsection{Serving and Protocol Measurements}
\label{app:workflow-scalability}
Cross-model means use three rounds of 128 requests per model and
scheme. Latency changes use the within-round \textsc{Vanilla}
baseline. CPU scaling uses one measurement per configuration.
The concurrency sweep uses one round per scheme and setting, with
16 requests at concurrency 1 and 128 at each of 8 and 32. The
concurrency-8 measurements reuse one Qwen3-1.7B round from the
cross-model benchmark.

\begin{tpapptable}{Cross-model serving at concurrency eight.}
	\label{tab:b4-cross-model}
	\begin{tabular*}{\columnwidth}{@{\extracolsep{\fill}}lrrrrr@{}}
		\toprule
		\textbf{Scheme} & \shortstack{Latency\\(s)}
		& \shortstack{$\Delta$ lat.\\(\%)} & \shortstack{Throughput\\(req/s)}
		& \shortstack{Online\\(ms/req)} & \shortstack{Output\\(tok/call)} \\
		\midrule
		\rowcolor{black!6}
		\multicolumn{6}{@{}l@{}}{\hspace{.4em}\textbf{Qwen3-1.7B}} \\
		V & \tpstat{19.19}{6.77} & 0 & \tpstat{0.465}{0.206} & \tpstat{1.30}{0.06} & \tpstat{39.97}{5.22} \\
		SL & \tpstat{19.96}{4.74} & \tpstat{+8.32}{19.14} & \tpstat{0.417}{0.111} & \tpstat{10.74}{0.47} & \tpstat{39.84}{5.37} \\
		SL+ & \tpstat{20.43}{4.98} & \tpstat{+10.75}{19.56} & \tpstat{0.407}{0.109} & \tpstat{58.17}{3.28} & \tpstat{39.93}{5.23} \\
		\textbf{TP} & \tpstat{18.78}{6.59} & \tpstat{-1.71}{6.88} & \tpstat{0.472}{0.199} & \tpstat{65.98}{12.87} & \tpstat{40.08}{5.37} \\
		\rowcolor{black!6}
		\multicolumn{6}{@{}l@{}}{\hspace{.4em}\textbf{Qwen3-4B}} \\
		V & \tpstat{15.80}{7.78} & 0 & \tpstat{0.577}{0.222} & \tpstat{1.17}{0.04} & \tpstat{39.06}{10.86} \\
		SL & \tpstat{15.85}{7.91} & \tpstat{+0.17}{4.14} & \tpstat{0.578}{0.224} & \tpstat{9.96}{0.69} & \tpstat{39.13}{11.04} \\
		SL+ & \tpstat{15.95}{6.82} & \tpstat{+2.87}{6.97} & \tpstat{0.555}{0.194} & \tpstat{57.63}{6.74} & \tpstat{39.07}{10.91} \\
		\textbf{TP} & \tpstat{15.28}{7.12} & \tpstat{-2.57}{3.10} & \tpstat{0.590}{0.216} & \tpstat{54.76}{5.31} & \tpstat{39.08}{10.98} \\
		\rowcolor{black!6}
		\multicolumn{6}{@{}l@{}}{\hspace{.4em}\textbf{Phi-3-mini (3.8B)}} \\
		V & \tpstat{105.38}{22.37} & 0 & \tpstat{0.078}{0.016} & \tpstat{1.25}{0.12} & \tpstat{256.00}{0.00} \\
		SL & \tpstat{83.28}{1.88} & \tpstat{-18.82}{15.35} & \tpstat{0.096}{0.002} & \tpstat{9.17}{0.29} & \tpstat{256.00}{0.00} \\
		SL+ & \tpstat{88.53}{5.72} & \tpstat{-13.74}{17.38} & \tpstat{0.091}{0.006} & \tpstat{57.83}{11.79} & \tpstat{256.00}{0.00} \\
		\textbf{TP} & \tpstat{96.42}{9.28} & \tpstat{-6.66}{14.10} & \tpstat{0.083}{0.008} & \tpstat{54.07}{6.73} & \tpstat{256.00}{0.00} \\
		\rowcolor{black!6}
		\multicolumn{6}{@{}l@{}}{\hspace{.4em}\textbf{DeepSeek-R1-Distill-Llama-8B}} \\
		V & \tpstat{72.62}{13.72} & 0 & \tpstat{0.112}{0.019} & \tpstat{1.22}{0.05} & \tpstat{255.95}{0.08} \\
		SL & \tpstat{69.23}{7.46} & \tpstat{-1.80}{25.08} & \tpstat{0.116}{0.012} & \tpstat{9.67}{0.43} & \tpstat{255.98}{0.03} \\
		SL+ & \tpstat{69.38}{6.67} & \tpstat{-2.23}{20.88} & \tpstat{0.116}{0.011} & \tpstat{56.86}{8.80} & \tpstat{255.98}{0.03} \\
		\textbf{TP} & \tpstat{76.70}{11.52} & \tpstat{+8.78}{30.63} & \tpstat{0.106}{0.015} & \tpstat{53.98}{7.07} & \tpstat{255.98}{0.03} \\
		\rowcolor{black!6}
		\multicolumn{6}{@{}l@{}}{\hspace{.4em}\textbf{Qwen3-14B}} \\
		V & \tpstat{35.77}{9.66} & 0 & \tpstat{0.236}{0.074} & \tpstat{1.15}{0.03} & \tpstat{68.15}{16.63} \\
		SL & \tpstat{37.28}{9.37} & \tpstat{+4.68}{4.72} & \tpstat{0.225}{0.066} & \tpstat{9.79}{0.55} & \tpstat{67.84}{16.61} \\
		SL+ & \tpstat{36.00}{8.18} & \tpstat{+1.64}{5.97} & \tpstat{0.230}{0.058} & \tpstat{64.93}{9.97} & \tpstat{67.76}{16.39} \\
		\textbf{TP} & \tpstat{35.90}{9.27} & \tpstat{+0.62}{1.81} & \tpstat{0.234}{0.071} & \tpstat{54.52}{3.48} & \tpstat{68.14}{17.02} \\
		\bottomrule
	\end{tabular*}
	\par\smallskip
	{\scriptsize
		\justifying
		\noindent
		V, SL, SL+, and TP denote Vanilla, SignedLog, SignedLogPlus, and
		TokenProv. Online denotes non-LLM time.\par
	}
\end{tpapptable}

\begin{tpapptable}{Online non-LLM components (ms/request).}
	\label{tab:b2-component-breakdown}
	\begin{tabular*}{\columnwidth}{@{\extracolsep{\fill}}lrrrrr@{}}
		\toprule
		Model & Lifecycle & $H\!\to\!B$ & $B\!\to\!P$ & Commit. & Receipt \\
		\midrule
		Q1.7 & \tpstat{23.00}{4.59} & \tpstat{15.87}{3.76} & \tpstat{13.83}{2.64} & \tpstat{1.47}{0.08} & \tpstat{8.75}{1.63} \\
		Q4 & \tpstat{18.99}{1.83} & \tpstat{13.02}{1.58} & \tpstat{11.63}{1.08} & \tpstat{1.36}{0.13} & \tpstat{7.10}{0.59} \\
		Phi & \tpstat{19.05}{2.44} & \tpstat{13.10}{2.13} & \tpstat{11.28}{1.44} & \tpstat{1.41}{0.08} & \tpstat{6.45}{0.55} \\
		DS8 & \tpstat{19.12}{2.46} & \tpstat{12.84}{2.01} & \tpstat{11.30}{1.57} & \tpstat{1.35}{0.08} & \tpstat{6.56}{0.90} \\
		Q14 & \tpstat{19.14}{0.93} & \tpstat{13.04}{1.24} & \tpstat{11.42}{0.66} & \tpstat{1.37}{0.06} & \tpstat{6.82}{0.77} \\
		\bottomrule
	\end{tabular*}
\end{tpapptable}

\begin{tpapptable}{CPU protocol scaling (ms/request).}
	\label{tab:b3-protocol-scaling}
	\begin{tabular*}{\columnwidth}{@{\extracolsep{\fill}}rrrrr@{}}
		\toprule
		& \multicolumn{2}{c}{4 events/call}
		& \multicolumn{2}{c@{}}{8 events/call} \\
		\cmidrule(lr){2-3}\cmidrule(l){4-5}
		Calls & Build & Audit & Build & Audit \\
		\midrule
		1   & 3.48   & 1.82   & 2.16   & 2.22 \\
		8   & 9.52   & 10.18  & 12.59  & 13.76 \\
		32  & 35.94  & 38.87  & 74.94  & 71.81 \\
		128 & 160.79 & 154.50 & 190.25 & 212.52 \\
		512 & 566.98 & 620.65 & 764.82 & 864.69 \\
		\bottomrule
	\end{tabular*}
\end{tpapptable}

\begin{tpapptable}{Qwen3-1.7B concurrency sweep.}
	\label{tab:b3-concurrency}
	\begin{tabular*}{\columnwidth}{@{\extracolsep{\fill}}rlrrrrr@{}}
		\toprule
		Conc. & Scheme & \shortstack{Lat.\\(s)} & Req./s & Tok./s
		& \shortstack{Online\\(ms/req)} & \shortstack{Output\\(tok/call)} \\
		\midrule
		1  & V           & 6.893  & 0.145 & 65.51  & 1.18  & 56.45 \\
		& SL          & 6.995  & 0.143 & 64.56  & 3.57  & 56.45 \\
		& SL+         & 6.999  & 0.143 & 64.51  & 14.03 & 56.45 \\
		& \textbf{TP} & 7.033  & 0.142 & 64.20  & 13.33 & 56.45 \\
		8  & V           & 11.365 & 0.703 & 192.49 & 1.30  & 34.24 \\
		& SL          & 14.719 & 0.543 & 147.35 & 10.80 & 33.95 \\
		& SL+         & 14.997 & 0.529 & 144.64 & 58.24 & 34.16 \\
		& \textbf{TP} & 11.409 & 0.700 & 191.68 & 52.46 & 34.23 \\
		32 & V           & 67.262 & 0.472 & 245.10 & 1.00  & 64.91 \\
		& SL          & 89.193 & 0.356 & 184.74 & 11.17 & 64.91 \\
		& SL+         & 72.323 & 0.438 & 227.59 & 86.04 & 64.91 \\
		& \textbf{TP} & 59.350 & 0.535 & 277.84 & 85.01 & 64.90 \\
		\bottomrule
	\end{tabular*}
\end{tpapptable}

\subsection{Provider-Assurance Measurements}
\label{app:c-assurance}
\label{app:provider-assurance}

Two persistent CPU workers replay 1,024 retained calls from
128 Qwen3-1.7B requests in three rounds. The circuit checks the
accounting commitment, integer charges, and scoped request hash.
The host validates receipt-bound witnesses and linkage. Signing uses
1,024 receipts per mode in three paired ECDSA-P256 sessions with
host-side accounting.

\begin{tpapptable}{Proof and isolated-key signing costs.}
	\label{tab:c1-assurance-cost}
	\label{tab:c4-tee-signing}
	\begin{tabular*}{\columnwidth}{@{\extracolsep{\fill}}lr@{}}
		\toprule
		\rowcolor{black!6}
		\multicolumn{2}{@{}l@{}}{\hspace{.4em}\textbf{(a) Groth16 components (ms/call)}} \\
		Persistent workers & Cost \\
		\midrule
		Host witness and linkage checks & \tpstat{0.168}{0.018} \\
		Circuit input preparation       & \tpstat{0.234}{0.016} \\
		Circuit witness calculation     & \tpstat{35.542}{3.477} \\
		Groth16 proving                 & \tpstat{112.872}{4.549} \\
		Witness calculation + proving   & \tpstat{148.445}{7.644} \\
		Proof verification              & \tpstat{15.101}{1.550} \\
		Complete warm RPC               & \tpstat{164.525}{9.185} \\
		\bottomrule
	\end{tabular*}
	\par\smallskip
	\begin{tabular*}{\columnwidth}{@{\extracolsep{\fill}}lrrr@{}}
		\toprule
		\rowcolor{black!6}
		\multicolumn{4}{@{}l@{}}{\hspace{.4em}\textbf{(b) Isolated-key signing (ms/receipt unless noted)}} \\
		Operation & Software key & Isolated key & Paired $\Delta$ \\
		\midrule
		Signing API & \tpstat{0.534}{0.028} & \tpstat{0.763}{0.085} & \tpstat{0.229}{0.108} \\
		Hashing + signing & \tpstat{0.673}{0.038} & \tpstat{0.887}{0.095} & \tpstat{0.213}{0.127} \\
		External verification & \tpstat{0.298}{0.021} & \tpstat{0.271}{0.025} & \tpstat{-0.026}{0.025} \\
		Key setup (ms/session) & \tpstat{2.135}{0.421} & \tpstat{7.070}{6.996} & \tpstat{4.934}{6.969} \\
		\bottomrule
	\end{tabular*}
\end{tpapptable}

Both modes persist witnesses in the foreground. I proves in
the foreground and D during audit. Total time sums both phases within
each round. Audit-rate costs sum operation wall times and divide by
all 100 eight-call requests. Elapsed time measures batch completion.

\begin{tpapptable}{Proof scheduling (ms/request).}
	\label{tab:c2-deferred-cost}
	\begin{tabular*}{\columnwidth}{@{\extracolsep{\fill}}rcrrrr@{}}
		\toprule
		Calls & Mode & Foreground & Proving & Audit & Total \\
		\midrule
		1 & I & \tpstat{151.59}{6.78} & \tpstat{148.07}{6.44} & \tpstat{14.98}{0.73} & \tpstat{166.57}{7.30} \\
		1 & D & \tpstat{2.57}{0.23} & \tpstat{141.01}{8.48} & \tpstat{156.17}{9.73} & \tpstat{158.73}{9.88} \\
		2 & I & \tpstat{287.55}{12.77} & \tpstat{282.06}{13.71} & \tpstat{29.21}{2.11} & \tpstat{316.76}{13.79} \\
		2 & D & \tpstat{3.03}{0.17} & \tpstat{290.83}{18.07} & \tpstat{322.12}{21.22} & \tpstat{325.15}{21.22} \\
		4 & I & \tpstat{593.34}{18.34} & \tpstat{587.42}{18.30} & \tpstat{59.29}{4.47} & \tpstat{652.63}{19.54} \\
		4 & D & \tpstat{4.92}{1.05} & \tpstat{576.59}{20.04} & \tpstat{639.04}{20.78} & \tpstat{643.96}{21.72} \\
		8 & I & \tpstat{1200.07}{60.99} & \tpstat{1188.31}{60.47} & \tpstat{117.47}{13.03} & \tpstat{1317.54}{74.00} \\
		8 & D & \tpstat{\textbf{8.65}}{2.48} & \tpstat{1201.46}{35.33} & \tpstat{1332.92}{39.68} & \tpstat{1341.56}{42.14} \\
		\bottomrule
	\end{tabular*}
\end{tpapptable}

\begin{tpapptable}{Selective auditing.}
	\label{tab:c3-audit-rate}
	\begin{tabular*}{\columnwidth}{@{\extracolsep{\fill}}rrrrrr@{}}
		\toprule
		Rate (\%) & Proofs & \shortstack{Proving\\(ms/req)}
		& \shortstack{Verify\\(ms/req)} & \shortstack{Combined\\(ms/req)}
		& \shortstack{Elapsed\\(s/batch)} \\
		\midrule
		1 & 8 & \tpstat{11.44}{0.38} & \tpstat{1.13}{0.09} & \tpstat{12.57}{0.47} & \tpstat{0.67}{0.05} \\
		5 & 40 & \tpstat{57.25}{0.63} & \tpstat{5.64}{0.23} & \tpstat{62.89}{0.53} & \tpstat{3.19}{0.04} \\
		\textbf{10} & 80 & \tpstat{112.55}{1.20} & \tpstat{10.98}{0.24} & \tpstat{\textbf{123.53}}{1.42} & \tpstat{\textbf{6.27}}{0.08} \\
		25 & 200 & \tpstat{286.97}{10.79} & \tpstat{29.15}{2.23} & \tpstat{316.11}{12.93} & \tpstat{15.95}{0.64} \\
		50 & 400 & \tpstat{565.33}{16.99} & \tpstat{55.68}{2.98} & \tpstat{621.01}{19.94} & \tpstat{31.29}{1.02} \\
		100 & 800 & \tpstat{1135.00}{33.13} & \tpstat{113.94}{8.11} & \tpstat{1248.94}{41.00} & \tpstat{62.88}{2.03} \\
		\bottomrule
	\end{tabular*}
\end{tpapptable}

\subsection{Evidence and Settlement Measurements}
\label{app:evidence-settlement}

Storage counts serialized bytes, and audit time covers in-memory
checks. Totals and their SDs use within-round sums.

\begin{tpapptable}{Evidence footprint and local settlement.}
	\label{tab:d-current-evidence-settlement}
	\begin{tabular*}{\columnwidth}{@{\extracolsep{\fill}}lrrrrr@{}}
		\toprule
		\rowcolor{black!6}
		\multicolumn{6}{@{}l@{}}{\hspace{.4em}\textbf{(a) Storage and non-LLM pipeline}} \\
		Model & \shortstack{Bundle\\(KiB/req)} & \shortstack{Persist\\(ms/req)}
		& \shortstack{Audit\\(ms/req)} & \shortstack{Remainder\\(ms/req)}
		& \shortstack{Non-LLM\\(ms/req)} \\
		\midrule
		Q1.7 & \tpstat{139.26}{0.13} & \tpstat{24.11}{4.93} & \tpstat{41.80}{8.38} & \tpstat{85.70}{16.34} & \tpstat{151.68}{29.21} \\
		Q4 & \tpstat{139.16}{0.11} & \tpstat{29.06}{8.87} & \tpstat{33.30}{3.87} & \tpstat{79.15}{13.50} & \tpstat{133.91}{18.81} \\
		Phi & \tpstat{139.41}{0.11} & \tpstat{23.17}{2.42} & \tpstat{36.05}{3.74} & \tpstat{76.83}{6.99} & \tpstat{130.91}{12.00} \\
		DS8 & \tpstat{140.59}{0.13} & \tpstat{19.78}{3.87} & \tpstat{34.38}{4.76} & \tpstat{71.49}{10.71} & \tpstat{125.47}{17.78} \\
		Q14 & \tpstat{139.23}{0.17} & \tpstat{21.16}{1.46} & \tpstat{34.90}{1.82} & \tpstat{73.55}{3.77} & \tpstat{128.07}{7.24} \\
		\bottomrule
	\end{tabular*}
	\par\smallskip
	\begin{tabular*}{\columnwidth}{@{\extracolsep{\fill}}lrrrrrr@{}}
		\toprule
		\rowcolor{black!6}
		\multicolumn{7}{@{}l@{}}{\hspace{.4em}\textbf{(b) Local settlement (ms/request)}} \\
		Model & Closure & Merkle & Record & Construct. & Append & Total \\
		\midrule
		Q1.7 & \tpstat{1.78}{0.29} & \tpstat{0.70}{0.05} & \tpstat{3.76}{0.46} & \tpstat{6.24}{0.80} & \tpstat{8.01}{1.22} & \tpstat{14.25}{2.01} \\
		Q4 & \tpstat{1.66}{0.06} & \tpstat{0.66}{0.03} & \tpstat{3.27}{0.11} & \tpstat{5.59}{0.17} & \tpstat{6.88}{0.44} & \tpstat{12.47}{0.60} \\
		Phi & \tpstat{1.53}{0.17} & \tpstat{0.70}{0.02} & \tpstat{2.98}{0.04} & \tpstat{5.21}{0.16} & \tpstat{7.60}{1.24} & \tpstat{12.81}{1.17} \\
		DS8 & \tpstat{1.90}{0.30} & \tpstat{0.67}{0.04} & \tpstat{3.20}{0.32} & \tpstat{5.77}{0.63} & \tpstat{6.94}{0.94} & \tpstat{12.71}{1.56} \\
		Q14 & \tpstat{1.84}{0.17} & \tpstat{0.68}{0.04} & \tpstat{3.26}{0.29} & \tpstat{5.78}{0.32} & \tpstat{6.89}{0.46} & \tpstat{12.67}{0.76} \\
		\bottomrule
	\end{tabular*}
	\par\smallskip
	{\scriptsize
		\justifying
		\noindent
		Q1.7, Q4, and Q14 denote the Qwen3 models. Phi and DS8 denote
		Phi-3-mini and DeepSeek-R1-Distill-Llama-8B.
		Local ledger size is fixed at 1.318\,KiB/request.\par}
\end{tpapptable}

\begin{tpapptable}{Storage comparison (KiB/request).}
	\label{tab:d-fulltrace-storage}
	\begin{tabular*}{\columnwidth}{@{\extracolsep{\fill}}lrrrrrrr@{}}
		\toprule
		Calls & 1 & 2 & 4 & 8 & 16 & 32 & 64 \\
		\midrule
		TokenProv  & 7.00 & 11.48 & 20.45 & 38.39 & 74.44  & 146.44 & 290.46 \\
		FullTrace  & 9.05 & 15.38 & 28.04 & 53.36 & 104.23 & 205.86 & 409.13 \\
		Saved (\%) & 22.69 & 25.35 & 27.07 & 28.05 & 28.59 & 28.87 & 29.01 \\
		\bottomrule
	\end{tabular*}
	\par\smallskip
	{\scriptsize
		\justifying
		\noindent
		Five repetitions per configuration give identical sizes.\par}
\end{tpapptable}

\subsection{Security and Accountability Tests}
\label{app:e-security-accountability}

Each adaptive episode in Table~\ref{tab:e-security-accountability}(a)
holds the authorization and retained handoffs fixed and permits up to
16 audited queries. The cross-request binding test in
Table~\ref{tab:e-security-accountability}(b) uses distinct request
identifiers with a shared call label.

\begin{tpapptable}{Supplemental security and accountability tests.}
	\label{tab:e-security-supplement}
	
	\begin{tabular*}{\columnwidth}{@{\extracolsep{\fill}}ll@{\hspace{.65em}}ll@{}}
		\toprule
		\rowcolor{black!6}
		\multicolumn{4}{@{}l@{}}{\textbf{(a) Injection suite}} \\
		\multicolumn{2}{@{}l}{\textbf{SignedLog and TokenProv (9)}}
		& \multicolumn{2}{l@{}}{\textbf{TokenProv only (5)}} \\
		\cmidrule(r){1-2}\cmidrule(l){3-4}
		A1  & Unauthorized call
		& A2  & Unanchored call insertion \\
		A5  & Call relabeling
		& A3  & Observable-call omission \\
		A6  & $H\!\to\!B$ handoff tampering
		& A4  & Call replay \\
		A7  & $B\!\to\!P$ handoff tampering
		& A12 & Cross-call splicing \\
		A8  & Model/deployment substitution
		& A13 & Receipt omission/duplication \\
		A9  & Tokenizer/input substitution & & \\
		A10 & Cache/usage inflation        & & \\
		A11 & Incorrect price version      & & \\
		A14 & Aggregate-bill manipulation  & & \\
		\bottomrule
	\end{tabular*}
	
	\par\smallskip
	{\scriptsize Each injection class uses five seeds with
		100 trials per seed.\par}
	\smallskip
	
	\begin{tabular*}{\columnwidth}{@{\extracolsep{\fill}}lccc@{}}
		\toprule
		\rowcolor{black!6}
		\multicolumn{4}{@{}l@{}}{\textbf{(b) Collusion controls}} \\
		Colluding parties     & $H+B$       & $B+P$                  & $H+B+P$ \\
		\midrule
		Fixture             & Hidden call & Splicing           & Consistent trace \\
		Conflicting handoff & None        & Signed $H\!\to\!B$ & None \\
		Accepted trials     & 500/500     & \textbf{0/500}     & 500/500 \\
		Resp.\ interval     & ---         & $[B,P]$            & --- \\
		\bottomrule
	\end{tabular*}
	
\end{tpapptable}

\subsection{Additional Ablation Measurements}
\label{app:f-ablation}

For original decision $a(x)$ and first failure $f(x)$, the
override returns $a(x)\lor[f(x)=R]$. Each configuration uses
five runs with 100 warmups and 1,000 timed executions per run.
We time validation and field selection and serialize payloads
afterward. Increments pair runs.

\begin{tpapptable}{Cumulative and incremental validation costs.}
	\label{tab:f-cumulative-cost}
	\label{tab:f-paired-increments}
	\label{tab:f-run-means}
	\begin{tabular*}{\columnwidth}{@{\extracolsep{\fill}}lrrrr@{}}
		\toprule
		& Lifecycle & $+$ Handoff & $+$ Provider & $+$ Settlement \\
		\midrule
		\rowcolor{black!6}
		\multicolumn{5}{@{}l@{}}{\hspace{.4em}\textbf{(a) Aggregate costs}} \\
		Wall (ms) & \tpstat{6.026}{0.029} & \tpstat{11.452}{0.037} & \tpstat{14.361}{0.095} & \tpstat{16.704}{0.109} \\
		CPU (ms) & \tpstat{6.020}{0.031} & \tpstat{11.432}{0.030} & \tpstat{14.342}{0.092} & \tpstat{16.673}{0.103} \\
		$\Delta$ wall (ms) & \tpstat{6.025}{0.029} & \tpstat{5.426}{0.057} & \tpstat{2.909}{0.089} & \tpstat{2.343}{0.198} \\
		Payload (KiB) & 15.369 & 22.468 & 29.543 & 32.217 \\
		$\Delta$ payload (KiB) & 15.348 & 7.099 & 7.075 & 2.674 \\
		\midrule
		\rowcolor{black!6}
		\multicolumn{5}{@{}l@{}}{\hspace{.4em}\textbf{(b) Run-mean wall time (ms)}} \\
		1 & 6.027 & 11.480 & 14.318 & 16.721 \\
		2 & 5.994 & 11.442 & 14.260 & 16.865 \\
		3 & 6.042 & 11.472 & 14.379 & 16.618 \\
		4 & 6.064 & 11.391 & 14.335 & 16.729 \\
		5 & 6.002 & 11.473 & 14.512 & 16.589 \\
		CV (\%) & 0.478 & 0.321 & 0.661 & 0.652 \\
		\bottomrule
	\end{tabular*}
	\par\smallskip
	{\scriptsize
		\justifying
		\noindent
		The request-ID baseline uses 22\,B, \tpstat{0.47}{0.07}\,$\mu$s
		wall time and \tpstat{1.29}{0.22}\,$\mu$s CPU time.\par}
\end{tpapptable}

\endgroup

\end{document}